\documentclass{article}
\usepackage{amsmath}
\usepackage{bbold}
\usepackage{array}
\usepackage[utf8]{inputenc}
\usepackage{graphicx} 
\usepackage{pstricks}

\begin{document}

\title{Statistical distribution of quantum entanglement for a random
  bipartite state}

\date{}

\author{ Celine Nadal$^1$, Satya N. Majumdar$^1$, and Massimo Vergassola$^2$
\\ $^1$
\small{\textit{Laboratoire de Physique Th\'eorique et Mod\`eles Statistiques
 (UMR 8626 du
CNRS), }}\\
\small{\textit{Universit\'e Paris-Sud, B\^atiment 100, 
91405 Orsay Cedex, France}}\\
$^2$ \small{\textit{Institut Pasteur, CNRS URA 2171, F-75724 Paris 15, France }}}

\maketitle

\begin{abstract}
  We compute analytically the statistics of the Renyi and von Neumann
  entropies (standard measures of entanglement), for a random pure
  state in a large bipartite quantum system. The full probability
  distribution is computed by first mapping the problem to a random
  matrix model and then using a Coulomb gas method.  We identify three
  different regimes in the entropy distribution, which correspond to
  two phase transitions in the associated Coulomb gas. The two
  critical points correspond to sudden changes in the shape of the
  Coulomb charge density: the appearance of an integrable singularity
  at the origin for the first critical point, and the detachement of
  the rightmost charge (largest eigenvalue) from the sea of the other
  charges at the second critical point. Analytical results are
  verified by Monte Carlo numerical simulations. A short account
  of some of these results appeared recently in Phys. Rev. Lett.
  {\bf 104}, 110501 (2010).
\end{abstract}

\section{Introduction}

Entanglement plays a crucial role in quantum information and
computation as a measure of nonclassical correlations between parts of
a quantum system \cite{Chuang}.  The strength of those quantum
correlations is significant in highly entangled states, which are
involved and exploited in powerful communication and computational
tasks that are not possible classically.  Random pure states are of
special interest as their average entropy is close to its possible
maximum value \cite{Lubkin,don page}.  Taking a quantum state at
random also corresponds to assuming minimal prior knowledge about the
system \cite{Hall}. Random states can thus be seen as ``typical
states'' to which an arbitrary time-evolving quantum state may be
compared.  In addition, random states are useful in the context of
quantum chaotic or nonintegrable systems \cite{chaos,chaos2,anderson}.

There exist several measures for quantifying entanglement
\cite{Vidal}.  For a bipartite quantum system, the entropy (either the
von Neumann or the Renyi entropies) is a well-known measure of
entanglement.  For a multipartite system, the full distribution of
bipartite entanglement between two parts of the system has been
proposed as a measure of multipartite entanglement \cite{MultiEnt}.
The distribution of entropy in a bipartite system is thus generally
useful for characterizing entanglement properties of a random pure
state.

Statistical properties of observables such as the von Neumann entropy,
concurrence, purity or the minimum eigenvalue for random pure states
have been studied extensively~\cite{Lubkin,don
  page,parisi,Sommers,G-conc,giraud,znidaric,MinEv,book_min,Chen}.  In 
particular,
the average von Neumann entropy is known to be close to its maximal
value (for a large system).  In contrast, few studies have addressed
the full distribution of the entropy: only the distribution of the
purity for very small systems \cite{giraud} and partial information on
the Laplace transform of the purity distribution for large systems
\cite{parisi} have previously appeared in the literature.

Our purpose here is to compute the full distribution of the Renyi
entropies for a random pure state in a large bipartite quantum system.
In particular, we show that the common idea that a random pure state
is maximally entangled is not quite correct: while the average entropy
is indeed close to its maximal value \cite{Lubkin,don page}, the
probability of an almost maximally entangled state is in fact vanishingly
small. This statement requires to compute the full probability distribution of
the entropy, namely its large deviation tails, which is one of the
goals achieved in our paper.

The calculation of the Renyi entropies' distribution proceeds by
mapping the entanglement problem to an equivalent random matrix model,
which describes the statistical properties of the reduced density
matrix of a subsystem.  We can then use Coulomb gas methods borrowed
from random matrix theory. We identify three regimes in the
distribution of the entropy, as a direct consequence of two phase
transitions in the associated Coulomb gas problem. One of those
transitions is akin to a Bose-Einstein condensation, with one charge
of the Coulomb gas detaching from the sea of the other charges - or
equivalently one eigenvalue of the reduced density matrix becoming
much larger than the others.

This paper is a detailed version of a short letter that was
published recently~\cite{prlBE}.  It thus contains all explicit
formulas for our results and details about analytical proofs and
numerical simulations as well as new results, especially for the third
regime of the distribution (see below), the von Neumann entropy and
the maximal eigenvalue of the density matrix.
\\

The plan of the paper is as follows. In section \ref{sec:RandBip}, we
describe precisely our model of bipartite quantum system for the
direct product $\mathcal{H}_A\otimes \mathcal{H}_B$ of two Hilbert
spaces $\mathcal{H}_A$ and $\mathcal{H}_B$.  In section
\ref{sec:DistriEv}, we analyze the distribution of the eigenvalues
$\lambda_i$ of the reduced density matrices of the two subsystems.  In
particular, we compute the average density of eigenvalues and explain
the Coulomb gas method that we also use later for computing the
distribution of the Renyi entropy $S_q=\frac{1}{1-q}\ln \Sigma_q$
where $\Sigma_q=\sum_i\lambda_i^q$.  In section \ref{sec:PdfSigmaq},
we compute the full distribution of $\Sigma_q$ for a large system. We
find two phase transitions in the associated Coulomb gas, and thus
three regimes for the distribution of $\Sigma_q$.  In section
\ref{sec:Renyi}, using results from section \ref{sec:PdfSigmaq}, we
derive the distribution of the Renyi entropy $S_q$ as well as the
distribution of the von Neumann entropy (case $q\rightarrow 1$) and
the distribution of the largest eigenvalue ($q\rightarrow \infty$).
Finally in section \ref{sec:MonteCarlo}, we present results obtained
by Monte Carlo numerical simulations that we performed to test and
verify our analytical predictions.

\section{Random bipartite state}
\label{sec:RandBip}

In this section, we set the problem of bipartite entanglement for a
random pure state. We first describe a bipartite quantum system,
introduce then measures of entanglement (the von Neumann and Renyi
entropies) and give finally the precise definition of random pure
states.

\subsection{Entanglement in a bipartite quantum system}
\label{subsec:entanglement}

Let us consider a bipartite quantum system $A \otimes B$ composed of
two subsystems $A$ and $B$ of respective dimensions $N$ and $M$.  The
system is described by the product Hilbert space $\mathcal{H}_{AB}=
\mathcal{H}_{A} \otimes \mathcal{H}_{B}$ with
$N=\dim\left(\mathcal{H}_A\right)$ and
$M=\dim\left(\mathcal{H}_B\right)$.  Here, we shall be interested in
the limit where $N$ and $M$ are large and $c=\frac{N}{M}$ is fixed. We
shall take $N\leq M$, i.e. $c\leq 1$, so that $A$ and $B$ play the
role of the subsystem of interest and of the environment,
respectively.

Let $|\psi\rangle$ be a pure state of the full system.  Its density
matrix $\rho=|\psi\rangle \langle \psi|$ is a positive semi-definite
Hermitian matrix normalized as ${\rm Tr}\,\rho =\langle \psi|\psi
\rangle = 1$.  The density matrix can thus be diagonalized, its
eigenvalues are non-negative and their sum is unity. Subsystem $A$ is
described by its reduced density matrix $\rho_A={\rm
  Tr_B\left[\rho\right]}= \sum_{\alpha^B=1}^M \langle \alpha^B |\rho|
\alpha^B \rangle$, where $|\alpha^B\rangle$ is an orthonormal basis of
$\mathcal{H}_B$.  Similarly, $B$ is described by $\rho_B={\rm
 Tr_A\left[\rho\right]}$. It is easy to show that the reduced matrices 
$\rho_A$ and $\rho_B$     
share the same set of non-negative eigenvalues
$\{\lambda_1,...,\lambda_N\}$ with $\sum_{i=1}^N \lambda_i =1$.

Any pure state can be written as $|\psi\rangle=\sum_{i=1}^N
\sum_{\alpha=1}^M x_{i,\alpha}\: |i^A\rangle \otimes|\alpha^B\rangle$
where $|i^A\rangle \otimes|\alpha^B\rangle$ is a fixed orthonormal
basis of $\mathcal{H}_{AB}$. The singular value decomposition of the
matrix $x_{i,\alpha}$ permits to recast the previous expression
in the so-called Schmidt decomposition form:
\begin{equation}\label{eq:schmidt} 
|\psi\rangle=\sum_{i=1}^N \sqrt{\lambda_i}
|m_i^A\rangle \otimes|\mu_i^B\rangle
\end{equation}
where $|m_i^A\rangle$ and
$|\mu_i^B\rangle$ represent the eigenvectors of $\rho_A$ and $\rho_B$,
respectively, associated with the same eigenvalue $\lambda_i$. 

The representation (\ref{eq:schmidt}), namely the Schmidt number $n_S$
of strictly positive eigenvalues, is very useful for characterizing
the entanglement between subsystems $A$ and $B$. For example, let us
consider two limiting cases:

(i) If only one of the eigenvalues, say $\lambda_i$, is non zero then
$\lambda_i=1$, $n_S=1$ and the state of the full system
$|\psi\rangle=|m_i^A\rangle \otimes|\mu_i^B\rangle$ is a product
state, which is said to be separable. The system is unentangled.

(ii) If all the eigenvalues are equal ($\lambda_j=1/N$ for all $j$),
$n_S=N$ and $|\psi\rangle$ is a superposition of all product
states. The system is maximally entangled.
\\

A standard measure of entanglement between two subsystems $A$ and $B$
is the von Neumann entropy of either subsystem: $S_{\rm VN}=-{\rm
  Tr}\left[\rho_A \ln \rho_A \right]=-\sum_{i=1}^N \lambda_i \ln
\lambda_i$, which reaches its minimum $0$ when the system is
unentangled (situation (i) above) and its maximum $\ln N$ when the
system is maximally entangled (situation (ii)). Another useful measure
of entanglement is the Renyi entropy of order $q$ (for $q>0$):
\begin{equation}
\label{Renyi}
S_q=\frac{1}{1-q}\, \ln\left[\sum_{i=1}^N \lambda_i^q\right]\,,
\end{equation}
which also reaches its minimal value $0$ in situation (i) and its
maximal value $\ln N$ in situation (ii).  As one varies the parameter
$q$, the Renyi entropy interpolates between the von Neumann entropy
($q\rightarrow 1^+$) and $-\ln \lambda_{\rm max}$ ($q\rightarrow
\infty$) where $\lambda_{\rm max}$ is the largest eigenvalue of the
reduced density matrices.
\\

\subsection{Random pure states}
\label{subsec:randstate}

A pure state is called random when it is sampled according to the
uniform Haar measure, which is unitarily invariant. Specifically, a
random pure state is defined as $|\psi\rangle=\sum_{i=1}^N
\sum_{\alpha=1}^M x_{i,\alpha}\: |i^A\rangle \otimes|\alpha^B\rangle$,
where $|i^A\rangle \otimes|\alpha^B\rangle$ is a fixed orthonormal
basis of $\mathcal{H}_{AB}$ and where the variables
$\left\{x_{i,\alpha}\right\}$ are uniformly distributed among the sets
of $\left\{x_{i,\alpha}\right\}$ satisfying the constraint
$\sum_{i,\alpha} \left|x_{i,\alpha}\right|^2=1$ (normalization of
$|\psi\rangle$).  Equivalently, the probability density function (pdf)
of the $N\times M$ matrix $X$ with entries $x_{i,\alpha}$
can be written
\begin{equation}\label{pdfX}
  P(X)\propto \delta\left({\rm Tr}(XX^{\dagger})-1\right) \propto 
e^{-\frac{\beta}{2}{\rm Tr}(XX^{\dagger})}\;\delta({\rm Tr}(XX^{\dagger})-1)\,,
\end{equation}
with the second equality showing that the pdf can also be seen as a
Gaussian supplemented by the unit-trace constraint.

In the basis $|i^A\rangle$ of $\mathcal{H}_{A}$, the reduced density
matrix of subsystem $\mathcal{A}$ is simply given by
$\rho_A=XX^{\dagger}$.
In general, when $X$ is a $N\times M$ Gaussian random matrix, i.e.
$P(X) \propto e^{-\frac{\beta}{2}{\rm Tr}(XX^{\dagger})}$ (iid Gaussian
entries $x_{i,\alpha}$ that are real for a Dyson index $\beta=1$,
complex for $\beta=2$), the $N\times N$ matrix $XX^{\dagger}$ is
a Wishart matrix whose distribution of eigenvalues is
\cite{James}:
\begin{equation}
P_{Wishart}(\lambda_1,...,\lambda_N)\propto e^{-\frac{\beta}{2} \sum_i \lambda_i}\:
\prod_{i=1}^N \lambda_i^{\frac{\beta}{2} (M-N+1) -1}\: 
\prod_{i<j} |\lambda_i -\lambda_j|^{\beta}\,.
\label{Wishart}
\end{equation}
The Vandermonde determinant $\prod_{i<j} |\lambda_i
-\lambda_j|^{\beta}$ makes that the eigenvalues are strongly
correlated and they physically tend to repel each other.

The major difference between the matrix $\rho_A=XX^{\dagger}$ in the
quantum problem and a standard Wishart matrix stems from the unit
trace constraint ${\rm Tr}\left[\rho_A \right]=1$.  The constraint is
to be included in the distribution of the eigenvalues of $\rho_A$,
which is given \cite{don page,Sommers} by:
\begin{eqnarray}\label{jpdfEV}
P(\lambda_1,...,\lambda_N)= B_{M,N}\;
\delta\big( \sum_i \lambda_i-1 \big) \;
\prod_{i=1}^N \lambda_i^{\frac{\beta}{2} (M-N+1) -1}\: 
\prod_{i<j} |\lambda_i -\lambda_j|^{\beta}\,,
\end{eqnarray}
with $\beta=2$ (the $x_{i,\alpha}$ are complex) and the normalization
constant $B_{M,N}$ computed using Selberg's integrals \cite{Sommers}:
\begin{equation}\label{selberg}
B_{M,N}=\frac{\Gamma(M N \beta/2) \, \Gamma(1+\beta/2)^N
}{\prod_{j=0}^{N-1} \Gamma((M-j)\beta/2) \, \Gamma(1+(N-j)\beta/2)}\,.
\end{equation}
 The presence of a fixed trace constraint  (as in Eq. (\ref{jpdfEV}))
is known to have important consequences 
on the spectral properties of a matrix~\cite{Akemann,Laks}. We will see that
in the present context also, the fixed trace constraint does play
an important and crucial role. In particular, this constraint is
directly responsible for a Bose-Einstein type condensation transition
that will be discussed in the context of the 
probability
distribution of the entanglement entropy.

Since the eigenvalues $\lambda_i$ of $\rho_A$ are random variables for
a random pure state, any observable is a random variable as
well. Statistical properties of observables, namely of various
measures of entanglement such as the von Neumann entropy \cite{don
  page,foong}, $G$-concurrence \cite{G-conc}, purity
\cite{parisi,giraud} or minimum eigenvalue 
\cite{znidaric,MinEv,book_min,Chen}, have been
studied extensively.  In particular, Page \cite{don page} computed the
average von Neumann entropy in the limit $M\geq N\gg 1$: $\langle
S_{VN}\rangle \approx \ln N-\frac{N}{2 M}$.  He also conjectured its
value for finite $N$ and $M$ (which was proved later \cite{foong}).
In contrast, there have been few studies on the full distribution of
the entropy, except for the purity $\Sigma_2=\sum_i \lambda_i^2$ whose
distribution is known exactly for small $N$ ($2,3$ and $4$)
\cite{giraud}.  For large $N$, the Laplace transform of the purity
distribution (generating function of the cumulants) was studied
recently \cite{parisi} for positive values of the Laplace
variable. However, when inverted, the previous quantity provides only
partial information about the purity distribution.

Here, we compute analytically the full distribution of the Renyi
entropy $S_q$ (defined in Eq.~\eqref{Renyi}) or equivalently of
$\Sigma_q=\sum_{i=1}^N\lambda_i^q=\exp{\left[(1-q)S_q \right]}$, for
large $N$. As for the von Neumann entropy, the average value of the
Renyi entropies is close to their maximal value $\ln N$ (maximal
entanglement)\,: $\langle S_q \rangle \approx \ln N -\bar{z}(q)$,
where $\bar{z}(q)>0$ (for $q>0$) is independent of $N$ for large $N$.
For example, for $M\approx N$ and $q=2$, we have $\bar{z}(q=2)=\ln 2$.
However, we show below that the probability that $S_q$ approaches its
maximal value $\ln N$ is again very small.

\section{Distribution of the eigenvalues of $\rho_A$}
\label{sec:DistriEv}

The eigenvalues of the reduced density matrix $\rho_A$ are distributed
according to the law in Eq.~\eqref{jpdfEV}. Given this joint distribution,
the first natural object to study is the average spectral density 
$\rho_{N,M}(\lambda)= \frac{1}{N}\sum_{i=1}^N \left\langle\delta\left(
    \lambda-\lambda_i\right)\right\rangle$. 
This average density
$\rho_{N,M}\left( \lambda\right) d\lambda$ also gives the probability to find
an eigenvalue between $\lambda$ and $\lambda +d\lambda$ (the one-point
marginal of the joint distribution). 
For finite $(N,M)$, this average density was computed first for
$\beta=2$~\cite{SZ,KAT} and very recently for $\beta=1$~\cite{PV}.
However, these formulae involve rather complicated special
functions and taking the asymptotic large $N$, large $M$ limit
is nontrivial. Here we will take a complementary route which
is well suited to derive exactly the asymptotic limit. We will
take the limit $N\to \infty$, $M\to \infty$ but keeping their
ratio $0\le c=N/M\le 1$ fixed. For the spectral density, we will
henceforth use a shorthand notation  
$\rho_N(\lambda)= \rho_{N,N/c}(\lambda)$. We will show that
for large $N$ the limiting form of $\rho_N(\lambda)$ can be obtained
easily via using a Coulomb gas approach.

Due to the unit trace
constraint $\sum_{i=1}^N \lambda_i=1$, the typical amplitude of the
eigenvalues is $\lambda_{typ}\sim\frac{1}{N}$ for large $N$.  Since
$\lambda_{typ}\sim \frac{1}{N}$ (and $\rho_N$ is normalized to unity),
we expect (as will be proved below) that the average density
for large $N$ has a scaling form:
\begin{equation}
\label{averDens}
\rho_N\left( \lambda\right)\approx N \; \rho^*\left(\lambda N\right)\,.
\end{equation}

Using the Coulomb gas method explained in subsection
\ref{subsec:CoulombDens}, we find an exact expression for the rescaled
density $ \rho^*(x)$:
 \begin{eqnarray}
\label{averDensResc}
 \rho^*(x)=\frac{1}{2 \pi c x}\sqrt{x-L_1}\:\sqrt{L_2-x}\,,
\end{eqnarray}
where the right and left edges read $L_2=c\left(
  \sqrt{\frac{1}{c}}+1\right)^2$, $L_1=c\left(
  \sqrt{\frac{1}{c}}-1\right)^2$ and we recall that $c=N/M\leq
1$.

\begin{figure}
\includegraphics[width=10cm]{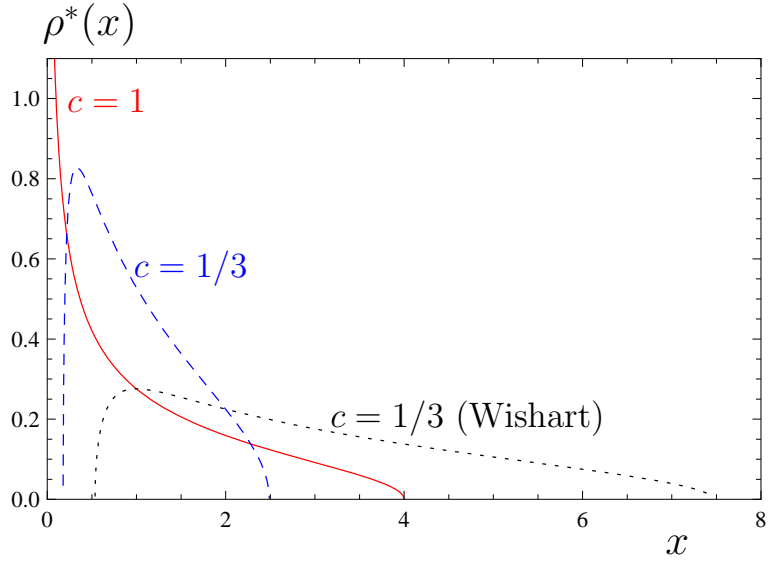}
\caption{The rescaled average density $ \rho^*(x)$ of the eigenvalues for
  the density matrix of a quantum subsystem.  The rescaled density is
  defined by $ \rho_N\left( \lambda\right)\approx N \;
  \rho^*\left(\lambda N\right)$ for large $N$ (see
  Eq.~\eqref{averDensResc}) and is plotted for $c=\frac{N}{M}=1$ (red
  solid line) and $c=1/3$ (blue dashed line).  The density is compared
  with the rescaled average density of Wishart eigenvalues (random
  matrix theory) : $\rho^*_W(x)$ defined by $\rho_N^W\left(
    \lambda\right)\approx \frac{1}{N}\: \;
  \rho^*_W\left(\frac{\lambda}{N}\right)$ (see Eq.~\eqref{averDensW})
  plotted for $c=\frac{N}{M}=1$ (red solid line) and $c=1/3$ (black
  dotted line). The different dependencies on $c$ for $ \rho^*(x)$ and
  $ \rho^*_W(x)$ make that, even after their different rescaling in
  $N$, the two distributions are equal only for $c=1$.
}\label{fig:averdens}
\end{figure}

For $c=1$ ($N\approx M$), $L_1=0$, $L_2=4$ and the rescaled density
reduces to:
 \begin{eqnarray}\label{averDensRescC1}
 \rho^*(x)=\frac{1}{2 \pi }\sqrt{\frac{4-x}{x}}\,.
\end{eqnarray}
In Fig.~\ref{fig:averdens}, plots of the rescaled density $
\rho^*(x)$ and comparisons to the shape of the rescaled density for a
standard Wishart matrix are shown for $c=1$ and $c=1/3$.

\subsection{Computation of the rescaled density: Coulomb gas method}
\label{subsec:CoulombDens}

The goal of this section is to prove Eqs.~\eqref{averDens} and
\eqref{averDensResc} for the average density of states.  The joint
distribution of the eigenvalues in Eq.~\eqref{jpdfEV} can be interepreted 
as a Boltzmann weight at inverse temperature $\beta$
\begin{equation}\label{jpdfBoltz}
P(\lambda_1,...,\lambda_N) \propto \exp\left\{ -\beta E \left[ \left\{ \lambda_i
    \right\} \right] \right\}\,,
\end{equation}
where the effective energy is given by
\begin{equation}
\label{jpdfEeff}
E \left[ \left\{ \lambda_i
  \right\} \right] = -\gamma \sum_{i=1}^N \ln \lambda_i -\sum_{i<j} \ln \left|\lambda_i-\lambda_j\right|
\;\;{\rm with}\;\; \sum_i \lambda_i=1\,.
\end{equation}
Here, $\gamma=\frac{M-N+1}{2}-\frac{1}{\beta}\approx N \frac{(1-c)}{2
  c}$ for large $N$. The logarithmic binary interactions correspond to
the Coulomb repulsion in $2$ dimensions. The eigenvalues can thus be
seen as charges of a $2D$ Coulomb gas, repelling each other
electrostatically. The charges are confined in the segment $1\geq
\lambda_i\geq 0$ for all $i$ and they are also subject to an external
logarithmic potential (with amplitude $\gamma$).

The mapping from random matrix eigenvalues to a Coulomb gas problem is
well-known in random matrix theory and has been recently used in a variety of
contexts that include the distribution of the extreme eigenvalues of
Gaussian and Wishart matrices~\cite{DM,vivo1,MV,KC}, purity partition
function in bipartite systems~\cite{parisi}, nonintersecting Brownian
interfaces~\cite{nadal1}, quantum transport in chaotic
cavities~\cite{vivo2}, information and communication
systems~\cite{kaz}, and the index distribution for Gaussian random
fields~\cite{BD,FW} and Gaussian matrices~\cite{nadal2}.  Here, we use
similar methods yet the problem is quite different due to the
constraint $\sum_i \lambda_i=1$. First, the scaling with $N$ (for
large $N$) differs from standard Wishart matrices. Indeed,
$\lambda_{typ}\sim 1/N$ in our problem of entanglement whereas
$\lambda_{typ}^W \sim N$ for a Wishart matrix. 
However, the effect of the constraint $\sum_i \lambda_i=1$ is 
not just the rescaling of standard Wishart results by a factor of $1/N^2$
as it may seem. It turns out that the constraint has more serious
consequences and leads to fundamentally different and new behavior
(including a condensation transition which is absent in Wishart matrices)
that we will demonstrate. 
\\

Configurations of the eigenvalues are characterized by the density
$\rho(\lambda,N)=N^{-1} \sum_{i=1}^N
\delta\left(\lambda-\lambda_i\right)$.  For large $N$, the eigenvalues
are expected to be close to each other and their typical amplitude is
$\lambda_{typ}\sim \frac{1}{N}$. We introduce then a rescaled variable
$x\sim O(1)$ as $x=\lambda N$.  The corresponding density is
$\rho(x)=N^{-1} \sum_{i=1}^N \delta\left(x-\lambda_i N\right)$, so
that $\rho(\lambda,N)=N\, \rho(\lambda N)=N\, \rho(x)$.

The effective energy in Eq.~\eqref{jpdfEeff} becomes in the continuous
limit (large $N$) a functional of the density $\rho$.  To the leading
order in $N$, the effective energy reads $ E \left[ \left\{ \lambda_i
  \right\} \right] =N^2 \, E \left[\rho \right]+O(N)$, where
\begin{eqnarray}
\label{jpdfEeffC}
 E\left[\rho \right]&=& -\left(\frac{1-c}{2 c}\right)
\int_0^{\infty} dx \:\rho(x) \ln x
-\frac{1}{2}\int_0^{\infty} \int_0^{\infty} dx dx' \:
\rho(x) \rho(x') \,\ln\left|x-x'\right|
\nonumber \\
&&+\,\mu_0 \left(\int_0^{\infty} dx \:\rho(x)-1\right) + \, \mu_1
 \left(\int_0^{\infty} dx \:x \,
\rho(x)-1\right)\,.
\end{eqnarray}
The Lagrange multipliers $\mu_0$ and $\mu_1$ enforce respectively the
constraints $\int \rho =1$ (normalization) and $\int dx\, x \,
\rho(x)=1$ (unit trace).

The joint distribution of the eigenvalues is given by the Boltzmann
weight $P(\lambda_1,...,\lambda_N)\propto \exp\left\{-\beta N^2
  E[\rho]+O(N)\right\}$ for large $N$. This distribution is highly
peaked around its most probable value $\rho^*$ which is thus also the
mean value of $\rho$: $\rho^*(x)=N^{-1} \sum_{i=1}^N \langle
\delta\left(x-\lambda_i N\right)\rangle $.  Hence, the average density
of states is the continuous density $\rho^*$ that
 minimizes the effective energy:
$\frac{\delta E}{\delta \rho}\Big|_{\rho=\rho^*}=0$.  From
Eq. \eqref{jpdfEeffC} we get the saddle point equation for $\rho^*$:
\begin{equation}\label{jpdfSaddle}
 \int_0^{\infty}dx'\:\rho^*(x') \,\ln\left|x-x'\right|=
\,\mu_0 + \, \mu_1\: x
 -\left(\frac{1-c}{2 c}\right)\, \ln x\,.
\end{equation}
Differentiating with respect to $x$ leads to the integral equation:
\begin{equation}
\label{jpdfSaddleTric}
\mathcal{P} \int_0^{\infty}dx'\:\frac{\rho^*(x')}{x-x'} =
\mu_1
 -\left(\frac{1-c}{2 c}\right)\frac{1}{x}\,,
\end{equation}
where $\mathcal{P}$ denotes the principal value.  This singular
integral equation can be solved by using a theorem due to
Tricomi~\cite{Tricomi} that states that if the solution
$\rho^*$ has a finite support $[L_1,L_2]$, then the finite Hilbert
transform defined by the equation $F(x)=\mathcal{P}
\int_{L_1}^{L_2}dx'\:\frac{\rho^*(x')}{x-x'}$ can be inverted as
\begin{equation}\label{Tricomi}
\rho^*(x)=\frac{1}{\pi \sqrt{x-L_1}\sqrt{L_2-x}}
\left[C-\mathcal{P}\int_{L_1}^{L_2}\frac{dx'}{\pi}
 \frac{\sqrt{x'-L_1}\sqrt{L_2-x'}}{x-x'}\,F(x')\right]\,,
\end{equation}
where the constant $C$ fixes the integral of $\rho^*$ via
$\int_{L_1}^{L_2}dx\, \rho^*(x)=C$.  

In Eq.~\eqref{jpdfSaddleTric}, $F(x)=\mu_1 -\left(\frac{1-c}{2
    c}\right)\frac{1}{x}$.  Physically, the average density is
expected to be smooth and thus to vanish at $L_1$ and $L_2$ (bounds of
its support): $\rho^*(L_1)=0=\rho^*(L_2)$.  These two constraints fix
the value of $L_1$ and $L_2$.  The other two constraints $\int
\rho^*=1$ and $\int x \rho^*=1$ give the value of the constant $C$ in
Eq.~\eqref{Tricomi} and the Lagrange multiplier $\mu_1$ in
Eq.~\eqref{jpdfSaddleTric}.  Finally, inserting the expression of
$\rho^*$ in Eq.~\eqref{jpdfSaddle} for a special value of $x$ (say
$x=L_2$) gives $\mu_0$. Imposing all these constraints, we finally
get:
\begin{equation}\label{averDensRescjpdf}
\rho^*(x)=\frac{1}{2 \pi c \, x}\sqrt{x-L_1}\,\sqrt{L_2-x}\,,
\end{equation}
with $L_{1,2}=c\left(1\mp\sqrt{\frac{1}{c}}\right)^2$ (where
$c=N/M$). We also find $C=\int \rho^*=1$, $\mu_1=1/\left(2 c\right)$
and $\mu_0=-\left(\frac{1+c}{2 c}\right)+2 \left(1-\frac{1}{c}\right)
\ln\left[1+\sqrt{c}\right] +\frac{\ln c}{2}$.  Finally, the average
density in the original variable $\lambda$ is given by $
\rho_N\left(\lambda\right)=N \:\rho^*\left(\lambda N\right)$,
where $\rho^*(x)$ is given in Eq.~\eqref{averDensRescjpdf}.

\subsection{Comparison with Wishart eigenvalues}
\label{subsec:AverDensWishart}

For Wishart matrices, it is known that the average density of the
eigenvalues is given, for large $N$ and fixed $c=N/M$, by the Mar\u
cenko-Pastur law~\cite{marcenko}:
\begin{equation}
\label{averDensW}
\rho_N^W\left( \lambda\right)\approx \frac{1}{N}\: \;
 \rho^*_W\left(\frac{\lambda}{N}\right)
\;\;{\rm with}\;\;
 \rho^*_W(x)=\frac{1}{2 \pi  x}\sqrt{x-L_1^W}\:\sqrt{L_2^W-x}\,,
\end{equation}
with the right and left edges given by $L_2^W=\left(1 +
  \sqrt{\frac{1}{c}}\right)^2$ and $L_1^W=\left(1 -
  \sqrt{\frac{1}{c}}\right)^2$.

As expected, the scaling with $N$ is different: $\lambda_{typ}^W\sim
N$ for a Wishart eigenvalue, whereas the unit trace constraint makes
that $\lambda_{typ }\sim 1/N$ for an eigenvalue of the quantum density
matrix $\rho_A$.

For $c=1$, the two edges $L_1^W=0$, $L_2^W=4$ and $ \rho^*_W(x)=
\rho^*(x)$.  However, for a general $c<1$ the rescaled densities are
not quite the same (even though they have the same shape): $
\rho^*_W(x)=c \: \rho^*(x c)$.  Figure \ref{fig:averdens} shows a
comparative plot of $ \rho^*_W(x)$ and $\rho^*(x)$ for $c=1$ and
$c=1/3$.

\section{Distribution of  $\Sigma_q=\sum_i \lambda_i^q$ for
$q>1$ and $c=1$}
\label{sec:PdfSigmaq}
This section is somewhat long as it contains the bulk of the details
of our calculations. Hence it is useful to start with a summary of the
main results obtained in subsections 4.1-4.3 as well as the main
picture that emerges out of these calculations. Readers not interested
in details can skip the subsections 4.1-4.3 and get the main
picture just from this summary.

In this section, we compute the full distribution of
$\Sigma_q=\sum_i\lambda_i^q$, and thus of the Renyi entropy
$S_q=\ln\left(\Sigma_q\right)/(1-q)$ for large $N$. We take for
simplicity $M\approx N$, i.e. $c=1$, but our method can be
easily extended to
$c<1$ as well. For simplicity, we will also restrict
ourselves to the case $q\ge 1$. However, our method is
also easily extendable to the case $0<q<1$.  
Since $\sum_i\lambda_i=1$ and $x\rightarrow x^q$ is
convex for $q>1$, we have $N^{1-q}\leq \Sigma_q \leq 1$ (or
equivalently $\ln N\geq S_q \geq 0$). The lower bound
$\Sigma_q=N^{1-q}$ corresponds to the maximally entangled case
(situation (ii) in subsection \ref{subsec:entanglement}), when
$\lambda_j=1/N$ for all $j$: the entropy is $S_q=\ln N$.  The upper
bound $\Sigma_q=1$ corresponds to the unentangled case (situation (i)
in subsection \ref{subsec:entanglement}) when only one of the
$\lambda_i$ is non zero (and thus equal to one): the entropy is zero.

\begin{figure}
\includegraphics[width=14cm]{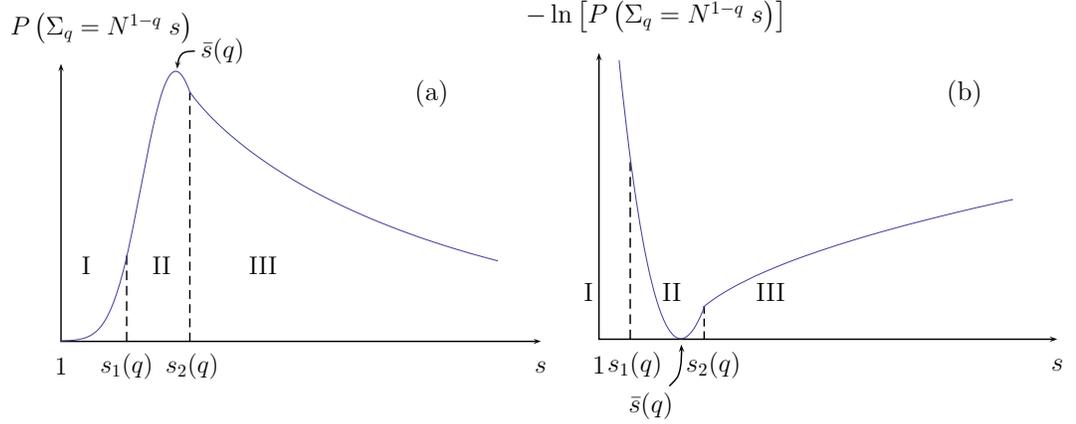}
\caption{Schematic distribution of $\Sigma_q =\sum_i
  \lambda_i^q=N^{1-q}\, s$ as a function of $s$ for (very) large $N$.
  Panel (a) shows the shape of the pdf of $\Sigma_q$, while (b) shows
  the shape of the rate function $-\ln P(\Sigma_q=N^{1-q}s)$.  Two
  critical points $s_1(q)$ and $s_2(q)$ separate three regimes {\bf
    I}, {\bf II} and {\bf III}, characterized by the different optimal
  densities shown in figure \ref{fig:schemadens}.  The maximally
  entangled state $s=1$ is at the extreme-left of the distribution,
  well spaced from the mean value $\bar{s}(q)$.}\label{fig:schemapdf}
\end{figure}

The scaling $\lambda_{typ}\sim 1/N$ implies that $\Sigma_q \sim
N^{1-q}$ for large $N$.  Let $s\equiv \Sigma_q \: N^{q-1}$ be the
rescaled variable $s\sim O(1)$.  In figure \ref{fig:schemapdf}, a
typical plot of the probability density function (pdf)
$P\left(\Sigma_q=N^{1-q}\,s \right)$ is shown: the distribution has a
Gaussian peak (centered on the mean value $s=\bar{s}(q)$) flanked on
both sides by non-Gaussian tails.  We show below that there are two critical
values $s=s_1(q)$ and $s=s_2(q)$ separating three regimes {\bf I}
($1\leq s< s_1(q)$), {\bf II} ($s_1(q)<s< s_2(q)$) and {\bf III}
($s_2(q)<s$).

At the first critical point $s_1(q)$, the distribution has a weak
singularity (discontinuity of the third derivative). At the second
critical point $s_2(q)$, a Bose-Einstein type condensation transition
occurs and the distribution changes shape abruptly (first derivative
is discontinuous in the limit $N\rightarrow +\infty$).  These changes
are a direct consequence of two phase transitions in the associated
Coulomb gas problem, more precisely in the shape of the optimal charge
density.  The schematic plot of the distribution of $\Sigma_q$ (for
large $N$) in Fig. \ref{fig:schemapdf} clearly shows the three regimes
{\bf I}, {\bf II} and {\bf III} and the discontinuity of the
derivative at $s=s_2$ (transition between {\bf II} and {\bf III}).
\\

\begin{figure}
\includegraphics[width=13cm]{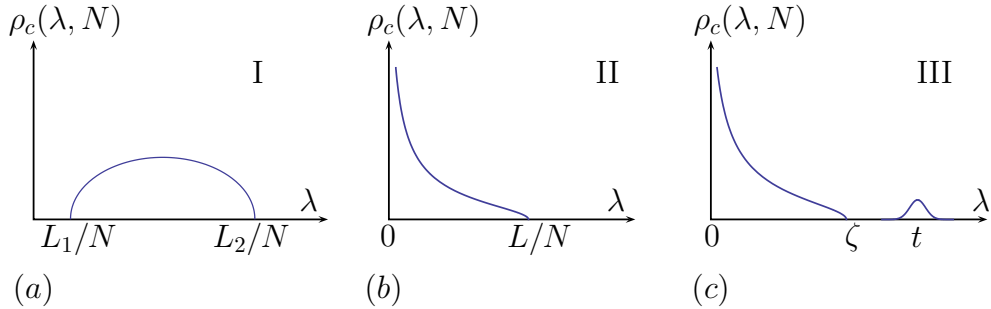}
\caption{Scheme of the optimal saddle point density $\rho_c$ of the
  eigenvalues (or, equivalently, of the Coulomb gas of charges) for (a)
  $1\leq s< s_1(q)$ (regime {\bf I}), (b) $s_1(q)<s< s_2(q)$ (regime
  {\bf II}) and (c) $s>s_2(q)$ (regime {\bf III}).  In regime {\bf
    III}, the maximal eigenvalue $\lambda_{\rm max}=t$ becomes much
  larger than the other eigenvalues, as shown by the isolated bump in
  (c).}\label{fig:schemadens}
\end{figure}

\begin{figure}
\includegraphics[width=13cm]{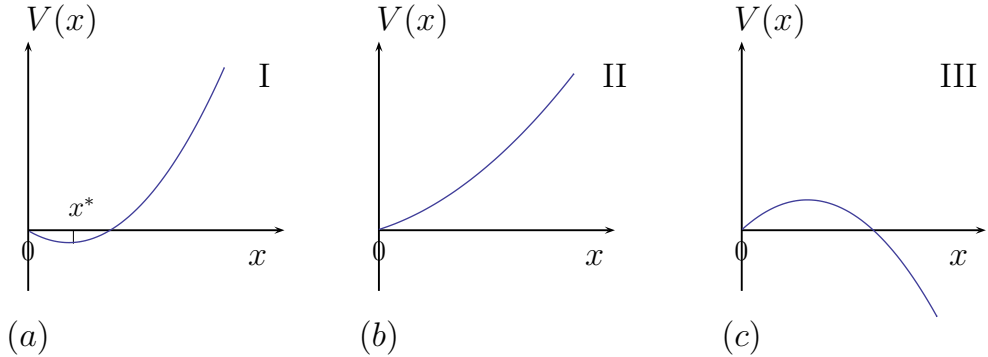}
\caption{Scheme of the effective potential $V(x)$ seen by the charges
  of the Coulomb gas (eigenvalues) for (a) $1\leq s< s_1(q)$ (regime
  {\bf I}), (b) $s_1(q)<s< s_2(q)$ (regime {\bf II}) and (c)
  $s>s_2(q)$ (regime {\bf III}).  In regimes {\bf I} and {\bf II}, the
  charges are confined close to the minimum of the effective
  potential.  In regime {\bf III}, the potential is not anymore
  bounded from below.  Therefore, one charge detaches from the sea of
  the other charges\,: the maximal eigenvalue becomes much larger than
  the other.}\label{fig:schemapot}
\end{figure}


More precisely, the probability density function of $\Sigma_q$ for
large $N$ and $q>1$ displays three different regimes:
\begin{equation}\label{PdfPurSummary}
P\left(\Sigma_q = N^{1-q}\, s \right)\approx
\left\{\begin{array}{ll}
\exp\left\{ -\beta N^2 \Phi_{I}(s)\right\}&{\rm for} \;\; 1\leq s< s_1(q)\,;\\
& \\
\exp\left\{-\beta N^2 \Phi_{II}(s)\right\}&{\rm for} \;\; s_1(q)<s< s_2(q)\,;\\
& \\
\exp\left\{-\beta N^{1+\frac{1}{q}}\; \Psi_{III}(s)\right\}&{\rm for} \;\; s>s_2(q)\,.
\end{array}
\right.
\end{equation}
The exact mathematical meaning of the ``$\approx$'' sign is a
logarithmic equivalence\,: $- \frac{\ln P\left(\Sigma_q = N^{1-q}\, s
  \right)}{\beta N^2} \longrightarrow \Phi_I(s)$ as $N\rightarrow
\infty$ with fixed $s \in [1,s_1(q)[$ (resp. $\Phi_{II}$ for fixed
$s\in ]s_1(q),s_2(q)[$) and $- \frac{\ln P\left(\Sigma_q = N^{1-q}\, s
  \right)}{\beta N^{1+1/q}} \longrightarrow \Psi_{III}(s)$ as
$N\rightarrow \infty$ with fixed $s > s_2(q)$.  The rate functions
$\Phi_I$, $\Phi_{II}$ and $\Psi_{III}$ (as well as $s_1$ and $s_2$)
are independent of $N$ - but they depend on the parameter $q$.
Explicit expressions of the functions $\Phi_I$ and $\Phi_{II}$ are
given in Eqs.~\eqref{PhiIq2} and \eqref{PhiIIq2} for $q=2$, and
in Eq.  \eqref{PhiII} for a general $q>1$; an explicit expression
of $\Psi_{III}$ is given in Eq.~\eqref{PsiIIIgen} for a general $q>1$
(and in Eq.~\eqref{PsiIIIq2} for $q=2$).  As shown in
figures~\ref{fig:N50Meth1} and \ref{fig:N1000Meth2} (resp. for $N=50$
and $N=1000$), we also did some Monte Carlo simulations (as explained
in section~\ref{sec:MonteCarlo}) and found that our analytical
predictions agree very well with the numerical data.

Regime {\bf II} includes the mean value $\langle \Sigma_q\rangle
\approx N^{1-q} \bar{s}(q)$, i.e. $s_1(q)<\bar{s}(q)\leq s_2(q)$ for
every $q$.  The mean value is explicitely given by:
\begin{equation}\label{SigmaQMean}
\langle \Sigma_q\rangle \approx N^{1-q} \bar{s}(q)\;\;\;
{\rm with} \;\;\;
\bar{s}(q)=
\frac{\Gamma(q+1/2)}{\sqrt{\pi } \Gamma(q+2)}\,
4^q\,.
\end{equation}
For large $N$, the distribution of $\Sigma_q$ given in
Eq. \eqref{PdfPurSummary} is highly peaked around its average (because
of the factor $N^2$ in regime {\bf II}): the average value of
$\Sigma_q$ coincides then with the most probable value,
i.e. $\bar{s}(q)$ is the minimum of $\Phi_{II}(s)$.  The quadratic
behaviour of $\Phi_{II}(s)$ around this minimum gives the Gaussian
behaviour of the distribution of $\Sigma_q$ around its average (and
thus gives the variance of $\Sigma_q$).  We get:
\begin{equation}\label{SigmaQGauss}
P\left(\Sigma_q =N^{1-q} s\right)\approx
\exp\left\{-\beta N^2 \frac{(s-\bar{s}(q))^2}{2 \sigma_q^2}\right\}\;\;
\textrm{for $s$ close to $\bar{s}(q)$}\,.
\end{equation}
Therefore, the variance of $\Sigma_q$ is given by:
\begin{equation}\label{SigmaQVar}
{\rm Var}\, \Sigma_q = \langle \Sigma_q^2 \rangle-\langle \Sigma_q\rangle^2
 \approx  \frac{\sigma_q^2}{\beta N^{2 q}}\;\;
{\rm with}\;\;\sigma_q^2=\frac{4^{2 q}}{2 \pi}q(q-1)^2
\frac{\Gamma(q+1/2)^2}{\Gamma(q+2)^2}\,.
\end{equation}
The distribution has a Gaussian peak flanked by non-Gaussian tails
described by the rate functions $\Phi_I$ (left tail) and $\Psi_{III}$
(right tail). Conversely, the rate function $\Phi_{II}$ describes the
middle part of the distribution, which includes the Gaussian behaviour
in the neighbourhood of the average.

In the limit $N\rightarrow \infty$, $s_1(q)$ and $s_2(q)$ do not
depend on $N$ and the second critical value $s_2(q)$ is actually equal
to the mean value $\bar{s}(q)$ of $s$:
\begin{equation}\label{CritPoints}
s_1(q)=\frac{\Gamma(q+3/2)}{\sqrt{\pi } \Gamma(q+2)}\,
\left(\frac{4(q+1)}{3 q}\right)^q
\;\;{\rm and}\;\; s_2(q)=\bar{s}(q)=
\frac{\Gamma(q+1/2)}{\sqrt{\pi } \Gamma(q+2)}\,
4^q\,.
\end{equation}
However, for a large but finite $N$, $s_2(q,N)$ actually depends on
$N$ and is given in Eq. \eqref{CritPointFiniteN} below.

The convergence in $N$ for the regimes {\bf I} and {\bf II} is very
fast\,: the agreement between numerical simulations and analytical
predictions in the limit $N\rightarrow \infty$ is very good already
for $N\simeq 50$.  However, the second transition, between regime {\bf
  II} and {\bf III}, is affected by finite-size effects, that remain
important even for $N\simeq O(10^3)$.  Their main effect is a shift in
the value of the critical point $s_2$.  The transition actually occurs
at a value $s_2(q,N)$ that depends on $N$, is a bit larger than
$\bar{s}(q)$ and tends slowly to $\bar{s}(q)$ as $N\rightarrow
\infty$.  More precisely, the second transition occurs at $s=s_2(q,N)$
with
\begin{equation}\label{CritPointFiniteN}
  s_2(q,N)\approx \bar{s}(q)+\frac{\left[\sqrt{q/2} \, (q-1) \, \bar{s}(q) \right]^{2
      q/(2 q-1)}}{
    N^{(q-1)/(2 q-1)}}\;\; \;\;
  \textrm{for large but finite $N$}\,.
\end{equation}
For example, for $q=2$, we have $\bar{s}(q=2)=2$ and
$s_2(q=2,N)\approx 2+\frac{2^{4/3}}{N^{1/3}} -\frac{2^{5/3}\ln N}{3
  N^{2/3}}$ for large but finite $N$.
\\

The extreme left of the distribution corresponds to maximally
entangled states: $s\rightarrow 1^+$ means that
$\sum_i\lambda_i^q=\Sigma_q \rightarrow N^{1-q}$, that is the case
where all the eigenvalues are equal and the state is maximally
entangled (situation (ii)).  As $s\rightarrow 1$, $\Phi_I(s)$ tends to
$+\infty$, thus the pdf $P(\Sigma_q=N^{1-q}s)$ tends rapidly towards
zero. For example, for $q=2$, we have $P(\Sigma_q=N^{1-q}s)\approx
(s-1)^{\beta N^2/4}$ as $s\rightarrow 1^+$.  This implies that the
probability of a maximally entangled configuration is very small (for
large $N$).

Similarly, the extreme right $s\rightarrow +\infty$ of the
distribution corresponds to weakly entangled states.  An unentangled
state has indeed only one non-zero eigenvalue, $\lambda_i$, thus
$S=\Sigma_q=1$ (situation (i)).  We can actually compute the
expression of the pdf for the scaling $\Sigma_q=S$ with $S\approx
O(1)$ ($S\gg s/N$) and $0<S<1$. For $q=2$, we get:
$P\left(\Sigma_2=S\right)\approx \left(1-\sqrt{S}\right)^{\beta
  N^2/2}$ for $N\rightarrow \infty$ with $S\approx O(1)$.  For
$S\rightarrow 1^-$, the pdf of $\Sigma_q$ is again tending very rapidly towards
zero: unentangled states are highly unlikely.
\\

The three regimes in the distribution of $\Sigma_q$
are actually a direct consequence of two phase transitions
in the associated Coulomb gas problem, as we show in this section.
We compute the probability density function $P(\Sigma_q=N^{1-q}\,s)$.
The charges of the associated Coulomb gas
see a different effective potential $V(x)$ depending on the value of $s$,
as shown by Fig.~\ref{fig:schemapot}:

$\bullet$ In regime {\bf I} ($1\leq s\leq s_1$), the potential $V(x)$
has a minimum at a positive $x$ and the charges accumulate near this
minimum: the optimal density $\rho_c(\lambda,N)$ describing the
charges has a finite support over $[L_1/N,L_2/N]$ and vanishes at
$L_1/N$ and $L_2/N$ (see Fig.~\ref{fig:schemadens}(a) and
\ref{fig:schemapot}(a)).

$\bullet$ In regime {\bf II} ($s_1<s\leq s_2$), the potential is
minimum at $x=0$, the charges accumulate close to the origin: the
optimal density $\rho_c(\lambda,N)$ describing the charges has a
finite support over $]0,L/N]$, vanishes at $L/N$ but diverges as
$1/\sqrt{\lambda}$ at the origin (see Fig.~\ref{fig:schemadens}(b) and
\ref{fig:schemapot}(b)).

$\bullet$ As $s$ exceeds $s_2$, the potential becomes unbounded from
below; the rightmost charge (maximal eigenvalue) suddenly jumps far
from the other eigenvalues: the charges are described in regime {\bf
  III} by a density with finite support $]0,\zeta]$ and a single
charge (maximal eigenvalue) well separated from the other charges: $t
\gg \zeta$ (see Fig.~\ref{fig:schemadens}(c) and \ref{fig:schemapot}(c)).

 \subsection{Computation of the pdf of $\Sigma_q$: associated Coulomb
   gas}

 In this subsection, we explain how we compute the pdf (probability
 density function) of $\Sigma_q$ using a Coulomb gas method.  The pdf
 of $\Sigma_q$ is by definition:
\begin{equation}\label{pdfSigmaq}
P(\Sigma_q,N)=\int
P(\lambda_1,...,\lambda_N) \:\: \delta\left(\sum_i\lambda_i^q -\Sigma_q \right)
\left(\prod_i d\lambda_i\right)\,.
\end{equation}
The joint pdf of the eigenvalues $P(\lambda_1,...,\lambda_N)$
is given in Eq.~\eqref{jpdfEV} and can be seen
as a Boltzmann weight at inverse temperature
 $\beta$, as in Eq.~\eqref{jpdfBoltz}:
\begin{equation}\label{jpdfBoltz1}
P(\lambda_1,...,\lambda_N) \propto \exp\left\{ -\beta E \left[ \left\{ \lambda_i
    \right\} \right] \right\}\,,
\end{equation}
where the energy $E \left[ \left\{ \lambda_i \right\} \right] =
-\gamma \sum_{i=1}^N \ln \lambda_i -\sum_{i<j} \ln
\left|\lambda_i-\lambda_j\right|$ (with $\sum_i \lambda_i=1$)
is the effective energy of a 2D Coulomb gas of charges.  For large
$N$, the effective energy is of order $E \sim O(N^2)$ (because of the
logarithmic interaction potential).  We can thus compute the multiple
integral in Eq. \eqref{pdfSigmaq} via the method of steepest descent:
for large $N$, the configuration of $\{\lambda_i\}$ which dominates
the integral is the one that minimizes the effective energy.

For Eq.~(\ref{pdfSigmaq}) we also have to take into account the
constraint $\sum_i\lambda_i^q=\Sigma_q$ (delta function in
Eq.~\eqref{pdfSigmaq}). This will be done by adding in the effective
energy a term $\mu'_2 \, \left(\sum_i\lambda_i^q-\Sigma_q\right)$
where $\mu'_2$ plays the role of a Lagrange multiplier.  Physically,
this corresponds to adding an external potential $\mu_2' \, \lambda^q$
for the charges.

For large $N$, the eigenvalues are expected to be close to each other
and the saddle point will be highly peaked, i.e. the most probable
value and the mean coincide.  We will thus assume that we can label
the $\lambda_i$ by a continuous average density of states
$\rho\left(\lambda,N \right)=N^{-1}\sum_i \langle
\delta(\lambda-\lambda_i)\rangle =N \,\rho(x)$ with
$\rho(x)=N^{-1}\sum_i \langle\delta(x-\lambda_i N)\rangle$ and
$x=\lambda N$.  However, we will see that this assumption is not
correct for large $\Sigma_q$ (large $s$): in the regime {\bf III}, the
maximal eigenvalue becomes much larger than the other eigenvalues. The
maximal eigenvalue should then be treated on its own and be
distinguished from the continuous average density.

Let us begin with the case where the eigenvalues can be described by
the density $\rho(x)$. Then the pdf of $\Sigma_q$ can be written as:
\begin{equation}\label{pdfSigmaqContinu}
P\left(\Sigma_q=N^{1-q} \, s,N\right) \propto \int \mathcal{D}\left[\rho\right]
\:\exp\left\{-\beta N^2 \, E_s\left[\rho\right]\right\}\,,
\end{equation}
where the effective energy $E_s\left[\rho\right]$ is given by
\begin{eqnarray}\label{EeffSq}
E_s\left[\rho\right]&=&-\frac{1}{2}\int_0^{\infty}\int_0^{\infty} dx dx' \:
\rho(x)\rho(x') \, \ln\left|x-x'\right|
+ \mu_0 \left(\int_0^{\infty} dx \: \rho(x)-1\right)\hspace{0.5cm}\nonumber\\
&&+\mu_1 \left(\int_0^{\infty} dx \:x\:  \rho(x)-1\right)
+\mu_2 \left(\int_0^{\infty} dx \: x^q\: \rho(x)-s\right) \,.
\end{eqnarray}
The Lagrange multipliers $\mu_0$, $\mu_1$ and $\mu_2$ enforce 
respectively the constraints
$\int \rho =1$ (normalization of the density),
$\sum_i\lambda_i=1$ (unit trace)
and $\sum_i \lambda_i^q=N^{1-q} \, s$ (delta function in Eq.\eqref{pdfSigmaq}).

For large $N$, the method of steepest descent gives:
\begin{equation}\label{SaddleEn}
P\left(\Sigma_q=N^{1-q} \, s,N\right)\propto \exp\left\{-\beta N^2 
E_s\left[\rho_c\right]\right\}\,,
\end{equation}
where $\rho_c$ minimizes the energy (saddle point):
\begin{equation}\label{SaddlePoint}
\frac{\delta E_s}{\delta \rho}\Big|_{\rho=\rho_c}=0 \,.
\end{equation}
The saddle point equation reads:
\begin{equation}\label{SaddleSq0}
\int_{0}^{\infty}dx'\, \rho_c(x') \ln\left|x-x'\right|
=\mu_0+\mu_1 x+\mu_2 x^q\equiv V(x) \,,
\end{equation}
with $V(x)$ acting as an effective external potential.
Differentiating with respect to $x$ gives:
\begin{equation}\label{SaddleSq}
\mathcal{P}\int_{0}^{\infty} dx'\, \frac{\rho_c(x')}{x-x'}
=\mu_1 +q \, \mu_2 x^{q-1}=V'(x) \,,
\end{equation}
where $\mathcal{P}$ denotes the Cauchy principal value.  The solution
for a finite support density $\rho_c$ is given again by Tricomi
formula as in Eq.~\eqref{Tricomi} and yields the answer for the
regimes {\bf I} and {\bf II}.

In these regimes, the pdf of $\Sigma_q$ is thus given by
$P\left(\Sigma_q=N^{1-q} \, s,N\right)\approx \exp\left\{-\beta N^2 \Phi(s)
\right\}$
where the rate function $\Phi(s)$ is equal to $E_s\left[\rho_c\right]$
up to an additive constant. 
More precisely, the normalized pdf reads:
\begin{equation}\label{pdfSigmaqContinuNorm}
P\left(\Sigma_q=N^{1-q} \, s,N\right)
 \approx \frac{\int \mathcal{D}\left[\rho\right]
\:\exp\left\{-\beta N^2 \, E_s\left[\rho\right]\right\}}{
\int \mathcal{D}\left[\rho\right]
\:\exp\left\{-\beta N^2 \, E\left[\rho\right]\right\}}\,,
\end{equation}
where $E_s\left[\rho\right]$ is given in Eq. \eqref{EeffSq} and
$E\left[\rho\right]$ is the effective energy associated to the joint
distribution of the eigenvalues (without further constraint), as given
in Eq. \eqref{jpdfEeffC} (we remind that $c=1$ in the present
section).  The steepest descent for both the numerator and denominator
gives:
\begin{equation}\label{SaddleEnNorm}
P\left(\Sigma_q=N^{1-q} \, s,N\right)\approx
\frac{\exp\left\{-\beta N^2 
E_s\left[\rho_c\right]\right\}}{
\exp\left\{-\beta N^2 
E\left[\rho^* \right]\right\}}
\approx \exp\left\{-\beta N^2 \Phi(s)\right\}\,,
\end{equation}
with
$\Phi(s)=E_s\left[\rho_c\right]-E\left[\rho^* \right]$ and 
where $\rho^*$ (resp. $\rho_c$) 
is the density that minimizes $E\left[\rho\right]$
(resp. $E_s\left[\rho\right]$).
The density $\rho^*(x)$ is thus simply the rescaled average density of states
given in Eq. \eqref{averDensRescC1} (for $c=1$).
Finally, we get 
\begin{equation}\label{PhiNormSq}
\Phi(s)=E_s\left[\rho_c\right]-E\left[\rho^* \right]=
E_s\left[\rho_c\right]-1/4 \,.
\end{equation}

\subsection{Regime {\bf I} and {\bf II}}

Regimes {\bf I} and {\bf II}
correspond to the case where the eigenvalues can be described by
a continous density $\rho(x)$, as explained above.
In this case, we have seen that the pdf of $\Sigma_q$
is given for large $N$ by
$P\left(\Sigma_q=N^{1-q} \, s,N\right)
\approx \exp\left\{-\beta N^2 \Phi(s)\right\}$.
In this section, we derive an explicit expression
for $\Phi(s)=\Phi_I(s)$ in regime {\bf I} 
ie for $1\leq s<s_1(q)$ (Eq. \eqref{PhiIq2}
in subsection \ref{subsec:regI} for $q=2$)
and $\Phi(s)=\Phi_{II}(s)$ in regime {\bf II}
ie for $s_1(q)<s<s_2(q)$ 
(Eq. \eqref{PhiIIq2} for $q=2$
and Eq. \eqref{PhiII} for a general $q>1$ in subsection \ref{subsec:regII}).

\subsubsection{Regime {\bf I}}\label{subsec:regI}

The solution of Eq. \eqref{SaddleSq} is a density with finite support
$[L_1,L_2]$ where $L_1\geq 0$. As the density is expected to be
smooth, we must have $\rho_c(L_2)=0$ and $\rho_c(L_1)=0$ at least for
$L_1>0$. As the eigenvalues $\lambda_i$ are nonnegative, another
possibility is that $L_1=0$ and $\rho_c(L_1) \neq 0$ -- this will be
regime {\bf II}.  The first case, i.e. with $L_1>0$ and
$\rho_c(L_1)=0$, defines the regime {\bf I} and is valid for $1\leq s
< s_1(q)$ with $s_1$ given in Eq. \eqref{CritPoints}, as we shall see
shortly.

In this subsection, we show that, for $1\leq s < s_1(q)$ (regime {\bf
  I}), $\mu_1<0$ and $\mu_2>0$, hence the effective potential $V(x)$
defined in Eq. \eqref{SaddleSq0} has a minimum at a nonzero $x$: at
$x=x^*=\left(\frac{-\mu_1}{q\, \mu_2}\right)^{\frac{1}{q-1}}\, >0$, as
shown by Fig.~\ref{fig:schemapot}(a).
  The charges concentrate around
this nonzero minimum.  Thus the density of charges $\rho_c$ is expected
to have a
finite support over $[L_1,L_2]$ with $L_1>0$ and to vanish at the
bounds $L_{1,2}$ (see Fig.~\ref{fig:schemadens}(a)).
\\

\textbf{A simple case: $q=2$}
\\

Let us begin with the case $q=2$, where we can find an explicit
expression for the density $\rho_c$ and the pdf of the purity
$\Sigma_2=\sum_i \lambda_i^2={\rm Tr} \left[\rho_A^2\right]$.

We find the solution of Eq. \eqref{SaddleSq} for $q=2$ by using
Tricomi formula with $F(x)=V'(x)$ (cf Eq.~\eqref{Tricomi}).  The
solution $\rho_c$ has a finite support $[L_1,L_2]$.  By imposing
$\rho_c(L_1)=0=\rho_c(L_2)$ (regime {\bf I}), we get:
\begin{equation}\label{RhocIq2proof}
\rho_c(x)=\frac{2 \mu_2}{\pi}\sqrt{x-L_1}\,\sqrt{L_2-x}\,.
\end{equation}
The optimal charge density is a semi-circle.
At this point, there are six unkown parameters: the constant $C$ in
Tricomi's formula; the bounds of the density support $L_1$ and $L_2$;
the Lagrange multipliers $\mu_0$, $\mu_1$ and $\mu_2$.  We also have
some constraints to enforce.  The two constraints
$\rho_c(L_1)=0=\rho_c(L_2)$, together with the three constraints $\int
\rho_c =1$, $\int x \rho_c=1$ and $\int x^2 \rho_c=s$ fix the value of
the five parameters $C$, $L_1$, $L_2$, $\mu_1$ and $\mu_2$. We get
$\mu_0$ by inserting the final expression of $\rho_c$ in
Eq. \eqref{SaddleSq0} for a special value of $x$, say $x=L_2$.

By imposing these constraints,
 we find $C=\int \rho_c=1$, $L_{1,2}=1 \mp 2\sqrt{s-1}$,
$\mu_1=-\frac{1}{2 (s-1)}$, $\mu_2=\frac{1}{4 (s-1)}$ and
$\mu_0=\frac{1}{2}\ln|s-1|+\frac{1}{4 (s-1)}-\frac{1}{2}$.
Therefore we have
\begin{equation}\label{RhocIq2}
\rho_c(x)=\frac{\sqrt{L_2 -x} \, \sqrt{x-L_1}}{2 \pi \, (s-1)}\,,
\end{equation}
with $L_{1,2}=1 \mp 2\sqrt{s-1}$.  This solution is valid for $L_1
>0$, i.e. for $s<5/4$.  Thus, regime {\bf I} corresponds to
$1\leq s<s_1(2)$ with $s_1(2)=5/4$.

In this regime, we have $\mu_1=-\frac{1}{2 (s-1)}<0$,
$\mu_2=\frac{1}{4 (s-1)}>0$, and the effective potential
$V(x)=\mu_0+\mu_1 x+\mu_2 x^2$ has a minimum for $x=x^*=1>0$. The
charges concentrate around this minimum: they form a semi-disk
centered at $x^*=1=(L_1+L_2)/2$.  The radius of the semi-disk $R=
2\sqrt{s-1}$ increases with $s$ till $L_1$ reaches its minimal
possible value $0$ (for $s=5/4$).

Finally we compute the saddle point energy.
 Using the
saddle point equation (Eq.~\eqref{SaddleSq0}), we get
$E_s\left[\rho_c\right]=-\frac{1}{2}\left(\mu_0+\mu_1+\mu_2 s \right)
=-\frac{1}{4}\ln\left(s-1\right)+\frac{1}{8}$, which gives
the expression of
$\Phi_I(s)=E_s\left[\rho_c\right]-E\left[\rho^*\right]
=E_s\left[\rho_c\right]-\frac{1}{4}$ (see Eq. \eqref{PhiNormSq}).
The distribution of the purity $\Sigma_2$ is thus given by:
\begin{equation}\label{pdfSigma2RegI}
P\left(\Sigma_2= s/N ,N\right)\propto \exp\left\{-\beta N^2 
\Phi_I(s) \right\}\,,
\end{equation} 
where the large deviation function $\Phi_I$ is explicitly given by:
\begin{equation}\label{PhiIq2}
\Phi_I(s)=-\frac{1}{4}\ln\left(s-1\right)-\frac{1}{8}\,.
\end{equation}
\\

\textbf{General case: $q>1$}
\\

The same qualitative behaviour holds for a general $q>1$: in the
regime {\bf I}, the effective potential $V(x)$ has a minimum at a
nonzero $x=x^*>0$, the charges accumulate around this minimum. The
density $\rho_c$ has a finite support $[L_1,L_2]$ with $L_1>0$ and
$\rho_c(L_1)=0=\rho_c(L_2)$.  This regime is valid for $1\leq s\leq
s_1(q)$. The value of the critical point is determined from the
analysis of regime {\bf II}: we show that regime {\bf II} is valid for
$s>s_1(q)$.  Unfortunately, we were not able to obtain explicit expressions 
for
$\rho_c$ and $\Phi_I$ in regime {\bf I} for general $q$ (the integral
in the Tricomi formula for a general
$q$ seems hard to compute analytically).

\begin{figure}
\includegraphics[width=10cm]{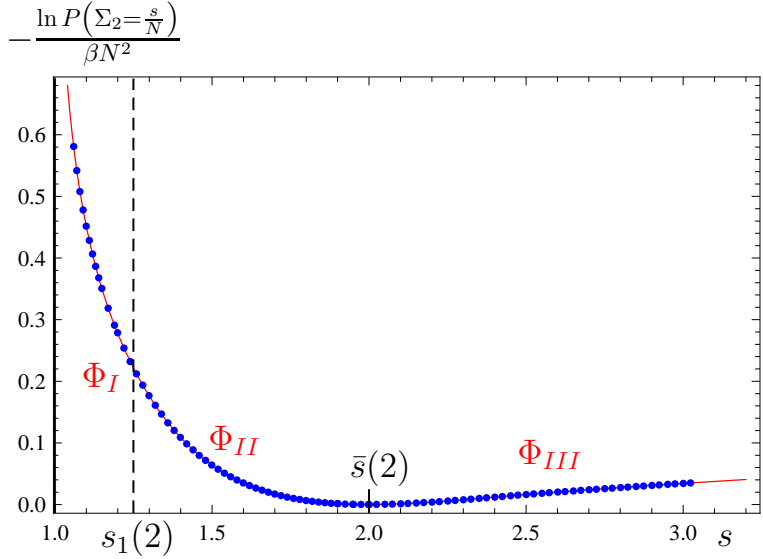}
\caption{Distribution of $\Sigma_2=\sum_i\lambda_i^2$\,: the figure
  shows the rate function $\Phi(s)=-\frac{\ln
    P\left(\Sigma_2=\frac{s}{N}\right)}{\beta N^2}$ plotted against
  $s$ for $N=50$.  Analytical predictions (red solid line) are
  compared with the results (blue points) of Monte Carlo numerical
  simulations (method 1, as explained in section
  \ref{sec:MonteCarlo}).  Our analytical predictions consist of three
  regimes.  For regimes {\bf I} ($1\leq s<5/4$) and {\bf II}
  ($5/4<s<2$), we have plotted the asymptotic expressions of the rate
  functions in the limit $N\rightarrow \infty$ given in
  Eqs.~\eqref{PhiIq2} and \eqref{PhiIIq2}.  For regime {\bf III}, we
  have plotted the analytical prediction for large but finite $N$,
  using for $\Phi_{III}(s,N)=\Phi(N,s/N)$ (see Eq.~\eqref{PhiIIIexpr})
  the complete expression of $E$ given in Eq.~\eqref{EnergyIII} and
  $\zeta$ and $t$ (numerical) solutions of Eq.~\eqref{eq:t} and
  \eqref{eq:zeta}.  Indeed, for $N=50$, finite-$N$ corrections to the
  asymptotic formula in Eq.~\eqref{PsiIIIq2} are important in regime
  {\bf III}\,: the curve of the dominant behavior in $N$ would not fit
  well the data and the complete expressions are needed. Note in
  particular that finite-$N$ effects make that the transition between
  {\bf II} and {\bf III} is regularized and appears to be
  smooth.}\label{fig:N50Meth1}
\end{figure}

\subsubsection{Regime {\bf II}}\label{subsec:regII}

As $s$ approaches $s_1(q)$ from below, the lower bound $L_1$ of the
density support tends to zero.  As the eigenvalues are non-negative,
$L_1$ cannot be negative. Hence, regime {\bf I} does not exist for
$s>s_1(q)$.  The critical value $s_1(q)$ is the onset of regime {\bf
  II}, where the density $\rho_c$ has a finite support $]0,L]$ and
vanishes only at the upper bound $L$ (see
Fig.~\ref{fig:schemadens}(b)).  We will see that regime {\bf II} is
valid for $s_1(q)\leq s \leq s_2(q,N)$ where $s_2(q,N)$ is given in
Eq.~\eqref{CritPointFiniteN}.

Within regime {\bf II} and for increasing $s$, $\mu_1$ increases and
becomes positive while $\mu_2$ remains positive. The effective
potential $V(x)=\mu_0+\mu_1 x+\mu_2 x^q$ has thus a minimum at a
smaller and smaller value $x=x^*$ that sticks to zero when $\mu_1$
becomes positive (see Fig.~\ref{fig:schemapot}(b)).  The charges
concentrate close to the origin.
\\

\textbf{A simple case: $q=2$}
\\

Let us begin with the simple case $q=2$. 
We find the solution of Eq. \eqref{SaddleSq} for $q=2$ by using again the
Tricomi formula with $F(x)=V'(x)$ (cf Eq. \eqref{Tricomi}).  We are
looking for a solution $\rho_c$ with finite support $[0,L]$.  After
imposing $\rho_c(L)=0$, we get:
\begin{equation}\label{RhocIIq2}
\rho_c(x)=\frac{1}{\pi}\sqrt{\frac{L-x}{x}}\left[A+B x\right]\,,
\end{equation}
with $A=\mu_1+\mu_2 L$ and $B=2 \mu_2$.

There are five unkown parameters: the arbitrary
constant $C$ in Tricomi's formula; the upper bound of the density
support $L$; the Lagrange multipliers $\mu_0$, $\mu_1$ and $\mu_2$.
We also have constraints to enforce.  The constraint $\rho_c(L)=0$
together with the three constraints $\int \rho_c =1$, $\int x
\rho_c=1$ and $\int x^2 \rho_c=s$ fix the value of the four parameters
$C$, $L$, $\mu_1$ and $\mu_2$. We get $\mu_0$ by inserting the final
expression of $\rho_c$ in Eq. \eqref{SaddleSq0} for a special value of
$x$, say $x=L$.

We find $C=\int \rho_c=1$, $\mu_1=8 (L-3)/L^2$, $\mu_2=4
(4-L)/L^3$ and 
$\mu_0=\ln\left(\frac{L}{4}\right)-\frac{1}{2}-\mu_1 \frac{L}{4}$.
  The upper bound of the support $L$ is solution of the
equation $L^2-12 L +16 s=0$. Hence $L=2(3 \pm \sqrt{9-4
  s})$. Physically the density $\rho_c(x)$ must remain positive for
$0<x<L$. It is not difficult to see that this determines $L$:
\begin{equation}
L=L(s)=2(3 -\sqrt{9-4 s})
\end{equation}
 The upper bound $L$
increases with $s$ and matches smoothly regime {\bf I}: $L=2=L_2$
at $s=s_1(2)=5/4$.  The  solution  of regime {\bf II}, 
exists as long as $s<9/4$.  However,
we shall see that there exists another solution for $s>2$ that is
energetically more favorable. This latter solution will yield regime
{\bf III}. The solution of regime {\bf II} is thus valid only for
$5/4<s<2$.

We have seen that $\mu_1=8 (L-3)/L^2$ and $\mu_2=4
(4-L)/L^3$.  According to the respective sign of $\mu_1$ and $\mu_2$,
we distinguish three phases for the effective potential
$V(x)=\mu_0+\mu_1 x+\mu_2 x^2$:
\begin{itemize}
\item $2\leq L<3$ (i.e. $5/4\leq s<27/16$): $\mu_1<0$ and $\mu_2>0$.
  The potential $V(x)$ has a minimum at a positive
  $x=x^*=\left(-\mu_1\right)/\left(2\, \mu_2\right)=L (3-L)/(4-L)$ (as
  in regime {\bf I}).  $x^*$ decreases when $L$ (or $s$) increases and
  reaches $0$ at $L=3$ (see Fig. 4 (a)).
\item $3< L <4$ (i.e. $27/16 <s<2$):  $\mu_1>0$ and $\mu_2>0$.
The potential is monotonic (increasing) on the real positive axis.
It has an absolute minimum at $x=0$ (see Fig. 4 (b)).
\item $L>4$ (i.e. $2<s\leq 9/4$): $\mu_1>0$ but $\mu_2<0$.  The
  potential is not anymore bounded from below.  It increases around
  the origin, reaches a maximum at
  $x=x^*=\left(\mu_1\right)/\left(-2\, \mu_2\right)=L (L-3)/(L-4)$ and
  decreases monotonically for $x>x^*$ to $-\infty$ (see Fig. 4 (c)). In this 
phase, the
  origin is a local minimum and the solution in Eq. \eqref{RhocIIq2}
  is metastable.  There is actually a second solution in this phase,
  where one eigenvalue splits off the sea of the other eigenvalues.
  This second solution becomes energetically more favorable at $s=s_2
  \approx 2 +\frac{2^{4/3}}{N^{1/3}}$.  The solution of regime {\bf
    II} in Eq.  \eqref{RhocIIq2} is thus valid only for $s<s_2$. For
  $s>s_2$, the second solution dominates: this is regime {\bf III}.
 \end{itemize}

 Finally, the distribution of the purity $\Sigma_2$ in regime {\bf II}
 is computed by the saddle point method:
\begin{equation}\label{PdfSigma2RegII}
P\left(\Sigma_2 =s/N ,N\right)\propto \exp\left\{-\beta N^2 
\Phi_{II}(s) \right\}\,,
\end{equation} 
where the large deviation function $\Phi_{II}=E_s\left[\rho_c\right]
-\frac{1}{4}=
 -\frac{1}{2}\left[\mu_1+\mu_2 s+\mu_0 \right]-\frac{1}{4}$
is  explicitely given by:
\begin{equation}
\label{PhiIIq2}
\Phi_{II}(s)=-\frac{1}{2}
\ln\left(\frac{L}{4}\right)+\frac{6}{L^2}-\frac{5}{L}+\frac{7}{8}\,,
\end{equation}
with $L=2\left(3-\sqrt{9-4 s}\right)$.  For large $N$, this solution
is valid for $s_1(2)<s\leq s_2(2,N)$ with $s_1(2)=5/4$ and $s_2(2,N)
\approx 2 +\frac{2^{4/3}}{N^{1/3}} \rightarrow 2$ as $N\rightarrow
+\infty$ (as we shall see).

  At $s=s_1=5/4$ (transition between regime {\bf I}
and {\bf II}), the rate
function $\Phi(s)$ has a weak nonanalyticity.  It is continuous,
$\Phi(5/4)=-\frac{1}{8}+\frac{\ln 2}{2}$, and even twice
differentiable: $\frac{d\Phi}{ds}\big|_{s=5/4}=-1$ and
$\frac{d^2\Phi}{ds^2}\big|_{s=5/4}=4$.  However, the third derivative
is discontinuous:
$\frac{d^3\Phi}{ds^3}\big|_{s=5/4^-}=\frac{d^3\Phi_I}{ds^3}\big|_{s=5/4}=-32$
but
$\frac{d^3\Phi}{ds^3}\big|_{s=5/4^+}=\frac{d^3\Phi_{II}}{ds^3}\big|_{s=5/4}=-16$.
The minimum of $\Phi$ is reached at $s=2$ within regime {\bf II},
 which gives the mean
value of the purity $\langle \Sigma_2 \rangle\approx 2/N$ (as the
distribution is highly peaked around its average for large $N$).

Figure \ref{fig:N50Meth1} compares our analytical predictions for
regimes {\bf I} and {\bf II} in Eq. \eqref{PhiIq2} and
\eqref{PhiIIq2} with numerical data (Monte Carlo simulations): the
agreement is very good already for $N=50$.
\\

\textbf{General case $q>1$}
\\

We find the solution $\rho_c$ with finite support $[0,L]$ of
Eq.~\eqref{SaddleSq} for $q>1$ by using again the Tricomi formula with
$F(x)=V'(x)$ (cf Eq. \eqref{Tricomi}).  After imposing $\rho_c(L)=0$,
we get the expression of the density:
\begin{eqnarray}\label{RhocIIgen}
  \rho_c=\frac{\mu_1}{\pi} \sqrt{\frac{L-x}{x}}+
  \frac{2\mu_2 q L^{q-1}}{\pi^{3/2}}  \frac{\Gamma
    \left(q+\frac{1}{2}\right)}{\Gamma(q)} 
  \sqrt{\frac{L-x}{x}} \:\,_2F_1\left(1,1-q,\frac{3}{2}, 1-\frac{x}{L}\right)\,,
\end{eqnarray}
where $\,_2F_1$ is a hypergeometric function
$\,_2F_1(a,b,c,z)=\sum_{n=0}^{\infty} \frac{(a)_n (b)_n}{(c)_n}
\frac{z^n}{n!}$, with $(a)_n=a (a+1)...(a+n-1)$ denoting the raising
factorial (Pochhammer symbol). 

  Exactly as for $q=2$, the constraints fix
the unknown parameters.  We obtain  the Lagrange multipliers
$\mu_1$, $\mu_2$  and $\mu_0$ as
functions of $L$: 
\begin{equation}\label{eq:mu12}
\mu_1=\frac{8 (1+q)}{(1-q) L^2} -\frac{4 q }{L (1-q)}\;\;\;\;\;
\textrm{and}\;\;\;\;\;
\mu_2 = \frac{ (1+q)}{(1-q) }\:
\frac{ \sqrt{\pi}\;  \Gamma(q)}{\Gamma(q+1/2)}\,\frac{L-4}{L^{q+1}}\,.
\end{equation}
and  
 $\mu_0=\ln\left(\frac{L}{4}\right)+\mu_1 \frac{L (1-q)}{2
  q}-\frac{1}{q}$.
 The upper bound $L$ (which is a function of $s$) is
given by the  
solution of the equation
\begin{equation}
\label{eq:L}
\left( \frac{1-q}{1+q} \right) L^q + 4 L^{q-1}=
\frac{2 \sqrt{\pi} \; \Gamma(q+1)}{\Gamma(q+1/2)}\, s\,.
\end{equation}
For $q=2$, we recover the simple expressions of the previous subsection.

The function $f: L\rightarrow \left( \frac{1-q}{1+q} \right) L^q + 4
L^{q-1}$ is increasing with $L$ for $0<L<L_0$ with $L_0=4(1+q)/q$, and
decreases for $L>L_0$.  It is thus maximal at $L=L_0$, which implies
that $s$ cannot be larger than $s_0=s(L=L_0)$ in this regime.  Hence,
regime {\bf II} is not valid for $s>s_0$, where
$s_0=s_0(q)=s(L=L_0)=\frac{\Gamma(q+1/2)}{2 \sqrt{\pi} \Gamma(q+2)}
\left(\frac{4 (1+q)}{q}\right)^q$.

Moreover, it can be shown that, for $L<L_0/3$ and for $L>L_0$, the
density $\rho_c(x)$ becomes negative for $x$ close to the bounds
(close to $0$ for $L<L_0/3$, close to $L$ for $L>L_0$).  This is not
physical.  Hence, $L$ must belong to the interval $[L_0/3,L_0]$.
Within this range, the function $f$ is monotonic and it increases with
$L$.  It can thus be inverted and gives $L$ as a single-valued function
of $s$: $L=L(s)$.  This range $[L_0/3,L_0]$ corresponds to $s_1(q)\leq s\leq
s_0(q)$, where $s_1(q)=s(L=L_0/3)$ and $s_0=s(L=L_0)$.

Therefore regime {\bf II} can exist only for $s_1(q)\leq s\leq
s_0(q)$, where $s_1(q)=\frac{\Gamma(q+3/2)}{ \sqrt{\pi} \Gamma(q+2)}
\left(\frac{4 (1+q)}{3 q}\right)^q$ and $s_0(q)=\frac{\Gamma(q+1/2)}{2
  \sqrt{\pi} \Gamma(q+2)} \left(\frac{4 (1+q)}{q}\right)^q$.  For
$q=2$, we recover $s_1(2)=5/4$ and $s_0(2)=9/4$.  However, as in the
$q=2$ case, this regime is not valid anymore for $s>s_2(q,N)$ given in
Eq.~\eqref{CritPointFiniteN}, where a second solution starts to
dominate (regime {\bf III}).

Finally, we compute the pdf of $\Sigma_q$ as a function of $L=L(s)$.
We get the pdf by the saddle point method:
\begin{equation}\label{PdfSigmaqRegII}
P\left(\Sigma_q =N^{1-q}\, s ,N\right)\propto \exp\left\{-\beta N^2 
\Phi_{II}(s) \right\}\,,
\end{equation} 
where the large deviation function $\Phi_{II}
=E_s\left[\rho_c\right]-\frac{1}{4}$
is explicitely given by:
\begin{equation}\label{PhiII}
  \Phi_{II}(s)=
  -\frac{1}{2} \ln \left(\frac{L}{4}\right)
  + \frac{4 (1+q)}{q L^2}-\frac{2 (1+2 q)}{q L}+\frac{3 q+1}{4 q}\,.
\end{equation}
The function $L=L(s)$ is the unique solution of Eq.~\eqref{eq:L}
within the range $s_1\leq s\leq s_2$.

Exactly as for $q=2$, the parameter $\mu_2$ (given in
Eq.~\eqref{eq:mu12}) is positive for $L<4$ ($s<\bar{s}(q)$) and
becomes negative for $L>4$ ($s>\bar{s}(q)$).  Hence, for all $q>1$ the
effective potential $V(x)=\mu_0+\mu_1 x+\mu_2 x^q$ becomes unbounded
from below when $L$ exceeds $4$.  The solution of regime {\bf II} is
thus metastable in the range $\bar{s}(q)<s<s_0(q)$ ($4<L<L_0$).
Indeed, exactly as for $q=2$, there exists a second solution for
$s>\bar{s}(q)$ that becomes energetically more favorable (lower
energy) for $s>s_2(q)$. This is the onset of regime {\bf III}. It
occurs at $s=s_2=\bar{s}$ for very large $N$, more precisely at
$s=s_2(q,N)\approx \bar{s}(q)+\left[\sqrt{q/2} \, (q-1) \, \bar{s}(q)
\right]^{2 q/(2 q-1)}/ N^{(q-1)/(2 q-1)}$ for large but finite $N$, as
we shall see.

As the distribution of $\Sigma_q$ is highly peaked for large $N$, its
mean value is given by the most probable value: $\langle \Sigma_q
\rangle=N^{1-q} \bar{s}(q)$ where $\bar{s}(q)$ minimizes $\Phi(s)$.
This minimum $s=\bar{s}(q)=\frac{\Gamma(q+1/2)}{\sqrt{\pi }
  \Gamma(q+2)}\, 4^q$ (or equivalently $L(\bar{s})=4$) is reached
within regime {\bf II} and $\Phi_{II}(\bar{s}(q))=0$.  For $s$ close
to $\bar{s}(q)$, $\Phi_{II}(s)\approx \frac{(s-\bar{s}(q))^2}{2
  \sigma_q^2}$ where $\sigma_q^2$ is given in Eq. \eqref{SigmaQVar}.
We conclude that the distribution of $\Sigma_q$ has a Gaussian
behaviour around its average, as shown in Eq.~\eqref{SigmaQGauss},
from which we can read the variance (see Eq.~\eqref{SigmaQVar}).  For
example, for $q=2$, we have $\sigma_2^2=4$ and ${\rm Var}
\Sigma_2\approx \frac{4}{\beta N^4}$.

\subsection{Regime {\bf III}}

As $s$ exceeds $\bar{s}(q)$, $\mu_2$ becomes negative and the
effective potential $V(x)=\mu_0+\mu_1 x+\mu_2 x^q$ is not anymore
bounded from below. The solution of regime {\bf II} becomes
metastable.  The minimum of the potential at the origin still exists,
as $V(x)$ increases for small $x$, but it is a local minimum: $V(x)$
reaches a maximum at $x=x^*>0$ and then decreases to $-\infty$ (see
Fig.~\ref{fig:schemapot}(c)).  Actually, for $s>\bar{s}(q)$, there
exists another solution where one charge splits off the sea of the
other $(N-1)$ charges that remain confined close to the origin (in the
local minimum of $V$). The maximal eigenvalue (charge) becomes much
larger than the other (see Fig.~\ref{fig:schemadens}(c)).  At some
point $s=s_2(q,N)$ very close to $\bar{s}(q)$ for large $N$, this
second solution becomes energetically more favorable than the solution
of regime {\bf II} : this is the onset of regime {\bf III}.  This
phase transition occurs at $s=s_2(q,N)$ given in
Eq. \eqref{s2qgen}. It is reminiscent of the real-space condensation
phenomenon observed in a class of lattice models for mass transport,
where a single lattice site carries a thermodynamically large
mass~\cite{MEZ}.

\subsubsection{Regime {\bf III}: summary of results}

We show in this section that there is an abrupt transition from regime
{\bf II} to {\bf III} at $s=s_2(q,N)$ where:
\begin{equation}\label{s2qgen}
  s_2(q,N)\approx  \bar{s}(q)+\frac{\left[\sqrt{q/2} \, 
(q-1) \, \bar{s}(q) \right]^{2
    q/(2 q-1)}}{N^{(q-1)/(2 q-1)}}\;
\;\textrm{for large $N$}\,.
\end{equation}
Here, $\bar{s}(q)$ the mean value of $s$ given in
Eq. \eqref{SigmaQMean}.  The maximal eigenvalue $t$ suddenly jumps
from a value $t\approx T/N$ very close to the upper edge
$\zeta$ of the sea of eigenvalues to a value $t\approx
\left[s-\bar{s}(q)\right]^{1/q}/N^{1-\frac{1}{q}}$ much larger than
the other eigenvalues ($t \gg \zeta$) (see Fig. 3 (c)).  This is clearly shown 
by the
good agreement between our predictions and numerical simulations in
Fig.~\ref{fig:max1} for $N=500$ and $N=1000$.  The consequence of this
phase transition in the Coulomb gas is an abrupt change in the
distribution of $\Sigma_q$.  More precisely, we show that for large
$N$\,:
\begin{equation}\label{PdfSigmaqRegIIILargeN}
P\left(\Sigma_q=N^{1-q}\, s,N \right)\approx \exp\left\{-\beta
  N^{1+\frac{1}{q}
}\:   \Psi_{III}(s)\right\}\;\;\; \textrm{for $s>s_2(q,N)$}\,, 
\end{equation}
where
\begin{equation}\label{PsiIIIgen}
\Psi_{III}(s)=\frac{\left[s-\bar{s}(q)\right]^{1/q}}{2}\,.
\end{equation}
The expression of the mean value $\bar{s}(q)$ is given in
Eq.~\eqref{SigmaQMean}.  For example, for $q=2$, this implies:
\begin{equation}\label{PsiIIIq2}
  P\left(\Sigma_2=\frac{s}{N},N \right)\approx \exp\left\{-\beta
    N^{\frac{3}{2}
    }\:   \Psi_{III}(s)\right\} \;\;
  {\rm with} \;\;\Psi_{III}(s)=\frac{\sqrt{s-2 }}{2}\,.
\end{equation}

The rate function $\Phi(N,s/N)$ defined by
\begin{equation}\label{RateFunQregIIandIII}
N^2\,\Phi(N,s/N)=\left\{\begin{array}{ll}
N^{2}\,  \Phi_{II}(s) &{\rm for} \;\; s<s_2\,,\\
N^{1+\frac{1}{q}}\, \Psi_{III}(s) &{\rm for} \;\; s>s_2\,,
\end{array}\right.
\end{equation}
is continuous but its derivative is discontinuous at $s=s_2$: for large
$N$
we have $ \frac{d\Phi}{ds}\big|_{s_2^+} \approx
\frac{d\Phi}{ds}\big|_{s_2^-} /(2 q)$.  At the transition point
$s=s_2$, there is also a change of concavity of the curve: the rate
function in regime {\bf II} is convex ($ \frac{d^2\Phi_{II}}{ds^2}>0$
for $s<s_2$) and has a minimum at $s=\bar{s}$, whereas the rate
function in regime {\bf III} is concave ($
\frac{d^2\Psi_{III}}{ds^2}<0$ for $s>s_2$).

Figure~\ref{fig:N1000Meth2} shows the transition from regime {\bf II}
to regime {\bf III} for $q=2$ and $N=1000$: analytical prediction for
large $N$ in Eq.~\eqref{PsiIIIq2} compare well with Monte Carlo
numerical simulations.

\subsubsection{New saddle point}

We want to describe the regime where a single charge (the maximal
eigenvalue) detaches from the continuum of the other charges.  The
assumption that all the eigenvalues are close to each other and can be
described by a continuous density of states does not hold anymore.  The
saddle point must be slightly revised.

We write $\lambda_{max}=t$ and label the remaining $(N-1)$ eigenvalues
by a continuous density $\rho(\lambda)=\frac{1}{N-1}\sum_{i \neq max}
\delta(\lambda-\lambda_i)$.  Physically, as the effective potential
has a local minimum at the origin $x=0$, we expect the optimal charge
density $\rho_c$ to have a finite support over $[0,\zeta]$ with
$\zeta<t$ and $\rho_c(\zeta)=0$: while one charge (the maximal
eigenvalue $t$) splits off the sea, the other charges (the sea) remain
confined close to the origin (in the local minimum of $V$, see Fig.
 \ref{fig:schemapot} (c)).

In this regime, we do not rescale the density (and the energy) by
assuming that $\lambda\sim 1/N$. We want indeed to compute the pdf of
$\Sigma_q=S$ for all $\bar{S}(q) \leq S \leq 1$, where
$\bar{S}(q)=N^{1-q}\, \bar{s}(q)$.  The effective energy is now a
function of both $t$ and $\rho$:
\begin{eqnarray}\label{EeffSqIII}
  E_S\left[\rho,t \right]=\!\!&-&\!\!\frac{(N-1)^2}{2}\int_0^{\zeta}\int_0^{\zeta} d\lambda d\lambda' \:
  \rho(\lambda)\rho(\lambda') \,
  \ln\left|\lambda-\lambda'\right|\nonumber\\ 
  &-&\!\!(N-1)\int_0^{\zeta}d\lambda \: \rho(\lambda) \, \ln\left|t-\lambda \right|
  +
  \mu_0 \left(\int_0^{\zeta} d\lambda \: \rho(\lambda)-1\right)
  \nonumber\\ 
  &+&\mu_1 \left((N-1) \int_0^{\zeta} d\lambda \:\lambda\:  \rho(\lambda)+t-1\right)
  \nonumber\\ 
  &+&\mu_2 \left((N-1) \int_0^{\zeta} d\lambda \: \lambda^q\: \rho(\lambda)+t^q-S\right)\,.
\end{eqnarray}

The dominating configuration is described by the optimal charge
density $\rho_c$ and the optimal value $t_c$ of $t=\lambda_{max}$ such
that:
\begin{equation}\label{SaddlePointSqIII}
\frac{\delta E_S}{\delta \rho}\Big|_{\rho=\rho_c,t=t_c}=0 \;\;{\rm  and}
 \;\;\frac{\partial E_S}{\partial t}\Big|_{\rho=\rho_c,t=t_c}=0\,.
\end{equation}
Taking into account the normalization, we have indeed for large $N$: $
P\left(\Sigma_q =S ,N\right)\approx \frac{ \int \mathcal{D} \rho \int
  dt \; e^{-\beta E_S\left[\rho,t\right]} }{ \int \mathcal{D} \rho
  \int dt \; e^{-\beta E\left[\rho,t\right]}} \approx
\exp\left\{-\beta \left( E_S\left[\rho_c,t_c \right]-
    E\left[\rho^*,t^* \right] \right) \right\} $, where
$E_S\left[\rho,t\right]$ is given in Eq. \eqref{EeffSqIII} and
$E\left[\rho,t\right]$ has the same expression as
$E_S\left[\rho,t\right]$ but without the last term (the constraint
$\sum_i \lambda_i^q=S$).  The pair $(\rho^*,t^*)$
(resp. $(\rho_c,t_c)$) minimizes $E\left[\rho,t\right]$
(resp. $E_S\left[\rho,t\right]$).  In fact, the normalization is given
by the saddle point energy evaluated at $S=\bar{S}$ (the mean value of
$S$):
$E\left[\rho^*,t^*\right]=E_S\left[\rho_c,t_c\right]\Big|_{S=\bar{S}}$
(with $\bar{S}=2/N$ for $q=2$).  We shall see that for large $N$, we
have:
\begin{equation}\label{NormIII}
E\left[\rho^*,t^*\right]=E_S\left[\rho_c,t_c\right]\Big|_{S=\bar{S}}\approx N^2\left(
\frac{\ln N}{2}+\frac{1}{4}\right)\,.
\end{equation}

Formally, by analogy with regimes {\bf I} and {\bf II}, we can write:
\begin{equation}\label{PdfSigmaqRegIII}
P\left(\Sigma_q =S ,N\right)\approx
 \exp\left\{-\beta N^2 \Phi(N,S)\right\}\,,
\end{equation}
where we define the rate function $\Phi$ as
\begin{equation}\label{PhiIIIexpr}
\Phi(N,S)=\left( E_S\left[\rho_c,t_c \right]-
    E\left[\rho^*,t^* \right] \right)/N^2\,.
\end{equation}
However, we shall see that the scaling of $\Phi$ with $N$ is different
in regime {\bf III} with respect to the regimes {\bf I} and {\bf II}.
In regimes {\bf I} and {\bf II}, $\Phi$ was independent of $N$ for
large $N$: $\Phi(N,s/N)\rightarrow \Phi_{I}(s)$
(resp. $\Phi_{II}(s)$).  In regime {\bf III}, we shall see that:
$\Phi(N,s/N)\approx \Psi_{III}(s)/N^{1-\frac{1}{q}}$ for large $N$.

For simplicity, we write $t$ instead of $t_c$ in the following.

\subsubsection{Case $q=2$}

Following the same steps as for regime {\bf II},
we find that the optimal charge density is explicitly given 
for $q=2$ by:
\begin{equation}\label{RhocIIIq2}
\rho_c(\lambda)=\frac{1}{\pi \, (N-1)}
\sqrt{\frac{\zeta-\lambda}{\lambda}}\left[
A+B \lambda +\frac{C}{t-\lambda} \right]\,,
\end{equation}
with
$A=\frac{4}{\zeta^2 }\left[N \zeta-2 
+2 \sqrt{t(t-\zeta)} \right]$ , 
$B=\frac{8}{\zeta^3 }\left[
4-N \zeta +\sqrt{\frac{t}{t-\zeta}}\,\left(3 \zeta-4 t\right)
\right]$ and 
$C=\sqrt{\frac{t}{t-\zeta}}$ ,
where $\zeta$ and $t=t_c$ satisfy:
\begin{eqnarray}
&(a)&\!\!\! 16 S+N \zeta^2 -12 \zeta -\sqrt{\frac{t}{t-\zeta}} \left(
16 t^2-20 t \zeta+5 \zeta^2 \right)=0 \label{eq:t}\,,\\
&(b)&\!\!\! \left(8 t^2-8 t \zeta +\zeta^2 \right)^2=8 (t-\zeta)
\sqrt{t (t-\zeta)} \left( 8 t -2 \zeta
-2 N t \zeta +N \zeta^2 \right)\,.\label{eq:zeta}
\end{eqnarray}
These equations can be solved numerically for every $\Sigma_2=S$.  We
can also find the solutions analytically for very large $N$.

For $S=\frac{s}{N}$ with $2<s<9/4$, there exist two solutions for the
pair $(\zeta,t)$. The first solution is of the form $t\approx \zeta$
with $\zeta \approx O(1/N)$. This is exactly (to leading order in $N$)
the solution of regime {\bf II} (see below, ``first solution'').
There is also a second solution, where $t \gg \zeta$: the maximal
eigenvalue becomes much larger than the other eigenvalues.  More
precisely, $\zeta \approx O(1/N)$ whereas $t\approx O(1/\sqrt{N})$ for
$S\approx O(1/N)$ (see below, ``second solution'').  We shall see that
the first solution (regime {\bf II}) is valid up to a value
$s=s_2\approx 2+\frac{2^{4/3}}{N^{1/3}}$ for large $N$, whereas the
solution with $t \gg \zeta$ starts to dominate for $s>s_2$ (its energy
becomes lower): this is regime {\bf III}.

For $S>\frac{9}{4 N}$ ($s>\frac{9}{4}$), there remains only one
solution (the second one), where $\zeta =L/N$ and $t\gg \zeta$.

Note that in both cases, for large $N$ (and for $\frac{2}{N}\leq S
<1$), the upper bound $\zeta$ remains of the order $\sim O(1/N)$. We
shall thus write $\zeta=\frac{L}{N}$ with $L\sim O(1)$.  On the other
hand, the maximal eigenvalue $t$ scales from $O(1/N)$ (as
$S\rightarrow 2/N$) to $O(1)$ (as $S\rightarrow 1^-$).
\\

Finally, we compute the saddle point energy as a function of
$\zeta=L/N$ and $t$. As finite-size effects (large but finite $N$) are
important in this regime, we keep all terms up to order $O(N)$ in the
saddle point energy, which gives:
\begin{eqnarray}
\label{EnergyIII}
E_S[\rho_c,t]=E(\zeta,t)\!\!\!\!&=&\!\!\!-\frac{(N-1)^2}{2}\ln\left[\frac{\zeta}{4}\right]
-2 N \ln\left[\frac{\sqrt{t}+\sqrt{t-\zeta}}{2}\right]
+\frac{1}{2}\ln\left[t (t-\zeta)\right]\nonumber \\
&&+\frac{9 N^2}{8}+\frac{6 (1+t^2)}{\zeta^2}-\frac{5  (N+t)}{\zeta}
+\frac{t}{8 (t-\zeta)} \nonumber \\
&&+ \sqrt{\frac{t}{t-\zeta}}\left[
-\frac{19 N}{4}-\frac{12 t}{\zeta^2}+\frac{11}{\zeta}+\frac{5 N t}{\zeta}
\right]\,,
\end{eqnarray}
where $\zeta=\zeta(s)$ and $t=t_c=t(s)$ are given by Eq. \eqref{eq:t}
and \eqref{eq:zeta}.

The rate function is thus given by
$\Phi(N,S)=\left( E_S\left[\rho_c,t \right]-
    E\left[\rho^*,t^* \right] \right)/N^2=\left( E\left[\zeta,t \right]-
    E\left[\rho^*,t^* \right] \right)/N^2$
with $ E\left[\zeta,t \right]$ given in Eq. \eqref{EnergyIII}.
\\

\textbf{Scaling $S=s/N$ with $s\sim O(1)$ : first solution
$t\approx\zeta$ with $\zeta \sim O(1/N)$ (regime {\bf II})}
\\

For $S=\frac{s}{N}$ with $s\sim O(1)$ for large $N$, the solution of
regime {\bf II} still exists as long as $s<9/4$ (where $9/4=s_0(2)$).
We recover this solution from the Eqs. ~\eqref{eq:t} and
\eqref{eq:zeta} with the scaling $t=\frac{T}{N}$ and
$\zeta=\frac{L}{N}$ with $T\approx L \sim O(1)$, i.e. the maximal
eigenvalue $t$ remains very close to the other eigenvalues
($t\approx\zeta$ for large $N$).

In this limit, equations \eqref{eq:t} and \eqref{eq:zeta}
indeed give:
\begin{eqnarray}
(a)\; 16 s+L^2-12 L&\approx&0\,, \label{eq:tLimII}\\
(b)\;\;\;\;\; (T-L)^{3/2}\;\;\;&\approx& \frac{L^{5/2}}{8 (6-L)}\frac{1}{N}\,.
\label{eq:zetaLimII}
\end{eqnarray}
Equation $(a)$ is the same as Eq. \eqref{eq:L} of regime
{\bf II}.
To leading order in $N$ (order $N^2$), Eq. \eqref{EnergyIII}
reduces to:
\begin{equation}\label{EnIIIlimII}
E_S[\rho_c,t]=E(L,t)=-\frac{N^2}{2}\ln\left(\frac{L}{4}\right)
+6 \frac{N^2}{L^2} -5 \frac{N^2}{L}+ N^2\left(
\frac{\ln N}{2}+\frac{9}{8}\right)\,.
\end{equation}
Therefore, using Eq. \eqref{NormIII}, we get
$\Phi(N,s/N)=\left(E_S[\rho_c,t]-E_S[\rho_c,t]\Big|_{s=2}\right)/N^2 =
\Phi(s)$ with $\Phi(s)=-\frac{1}{2}\ln\left(\frac{L}{4}\right)
+\frac{6}{L^2} - \frac{5}{L}+\frac{7}{8}=\Phi_{II}(s)$.  We recover
the expression in Eq. \eqref{PhiIIq2} of regime {\bf II}.

However, for $S=s/N>2/N$ there exists a second solution that becomes
energetically more favorable at some point $s_2\approx
2+\frac{2^{4/3}}{N^{1/3}}$.  Therefore regime {\bf II} is only valid
for $5/4<s<s_2$.
\\

\begin{figure}
\includegraphics[width=9cm]{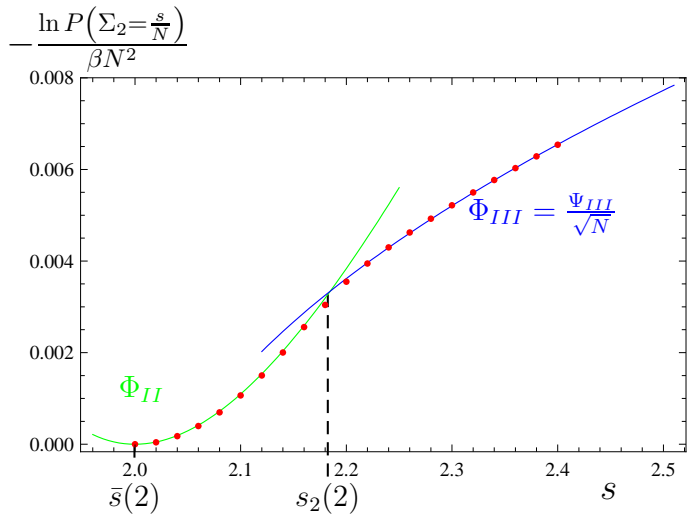}
\caption{Distribution of $\Sigma_2$\,: rate function $\Phi=-\ln
  P\left[\Sigma_2=s/N\right]/(\beta N^2)$ plotted against $s$ for
  $N=1000$. Analytical results (solid line) are compared with data
  (red points) of numerical simulations (Monte Carlo, method 2, see
  section \ref{sec:MonteCarlo}).  Analytical results here are the rate
  functions expected in the limit of very large $N$: $\Phi_{II}(s)$ in
  regime {\bf II} (green solid line, see Eq. \eqref{PhiIIq2}) and
  $\Phi(N,s/N)\approx \Psi_{III}(s)/\sqrt{N}$ in regime {\bf III}
  (blue solid line, see Eq. \eqref{PsiIIIq2}). The transition between
  regimes {\bf II} and {\bf III} is abrupt, we can see the
  discontinuity of the derivative of the rate function.  It occurs at
  $s_2(q=2,N) \approx 2+\frac{2^{4/3}}{N^{1/3}} -\frac{2^{5/3} \ln
    N}{3 N^{2/3}}\approx 2.18$ for $N=1000$.}\label{fig:N1000Meth2}
\end{figure}

\textbf{Scaling $S=s/N$ with $s\sim O(1)$ : second solution
$t\gg\zeta$ (regime {\bf III})}
\\

For $S=s/N$ with $s>2$, there exists a second solution where one
eigenvalue ($\lambda_{\rm max}=t$) becomes much larger than the
others\,: $t \gg \zeta$.  In this limit, Eq. \eqref{eq:t} and
\eqref{eq:zeta} give for large $N$:
\begin{eqnarray}\label{eq:tZetasol2}
 t\approx\frac{\sqrt{s-2}}{\sqrt{N}}
\;\; \;{\rm and}\;\;\;
\zeta \approx \frac{4}{N}
 \left[1+\frac{3-s}{\sqrt{s-2}}\;\frac{1}{\sqrt{N}}\right]\,.
\end{eqnarray}
For $S\rightarrow 1$, which implies $s\rightarrow \infty$ as
$N\rightarrow \infty$, we find $t\approx \sqrt{\frac{s}{N}}=\sqrt{S}$
and $\zeta \approx \frac{4}{N}\left(1-\sqrt{\frac{s}{N}}\right)
\approx\frac{4}{N}(1-t)\;$ as also recovered in Eq.~\eqref{RegIIIlimS1}.

We can expand the saddle point energy in Eq. \eqref{EnergyIII}
replacing $t$ and $\zeta$ by the expressions given in Eq. \eqref{eq:tZetasol2}
for large $N$. We obtain:
\begin{equation}\label{EnIIIsol2}
E_S\left[\rho_c,t\right]\approx 
\frac{\sqrt{s-2}}{2}\, N^{3/2}+N^2\left(\frac{\ln N}{2}+\frac{1}{4} \right)
-\frac{N}{2} \ln N +O(N)\,\,\,\,\textrm{for large $N$}\,.
\end{equation}
Finally, we get
$N^2 \Phi(N,s/N)=E_{s/N}\left[\rho_c,t\right]-
N^2\left(\frac{\ln N}{2}+\frac{1}{4} \right)\approx
\frac{\sqrt{s-2}}{2}\; N^{3/2}
-\frac{N}{2} \ln N +O(N)$ for large $N$ (see Eq. \eqref{NormIII}) and
 the pdf of $\Sigma_2$ is  thus given for large $N$ by:
\begin{equation}\label{PdfSigma2regIIILargeN}
P\left(\Sigma_2=\frac{s}{N},N \right)\approx e^{-\beta  N^{3/2} \Psi_{III}(s)}\,,
\end{equation}
where $N^{3/2}\Psi_{III}(s)=N^2 \Phi(N,s/N)$, that is
\begin{equation}\label{PsiIIIq2bis}
\Psi_{III}(s)=\frac{\sqrt{s-2}}{2}
-\frac{\ln N}{2 \sqrt{N}}  +O\left(\frac{1}{\sqrt{N}}\right)\approx
\frac{\sqrt{s-2}}{2}\;\;\; \textrm{for large $N$}\,.
\end{equation}
The rate function has a very different behaviour for large $N$ in
regime {\bf II} and {\bf III}.  In regime {\bf I} and {\bf II}, we
have $P\left(\Sigma_2=\frac{s}{N},N\right) \approx e^{-\beta N^{2}
  \Phi(s)}$, whereas in regime {\bf III} we have
$P\left(\Sigma_2=\frac{s}{N},N\right) \approx e^{-\beta N^{3/2}
  \Psi_{III}(s)}$.  For large but finite $N$ and for $s>2$ but very
close to $\bar{s}=2$, we have $ N^{3/2} \Psi_{III}(s)> N^{2}
\Phi_{II}(s)$.  Therefore the solution of regime {\bf II} dominates
close to $s=2$.  However, the solution of regime {\bf III} becomes
energetically more favorable at some point $s_2$ defined by $N^{3/2}
\Psi_{III}(s_2)= N^{2} \Phi_{II}(s_2)$, that is
\begin{equation}\label{s2q2}
  s_2 \approx 2+\frac{2^{4/3}}{N^{1/3}}
  -\frac{2^{5/3} \ln N}{3 N^{2/3}}
  \;\;\textrm{for large $N$}\,.
\end{equation}
At $s=s_2$, there is an abrupt transition from regime {\bf II} to {\bf
  III}. The maximal eigenvalue $t$ jumps from a value $t\approx
\frac{T}{N}$ with $T\sim O(1)$ and $t$ very close to $\zeta$ to a
value $t\approx \frac{\sqrt{s-2}}{\sqrt{N}}$ much larger than the
other eigenvalues ($t \gg \zeta$).  The rate function is continuous but
its derivative is discontinuous: $N^{2}
\frac{d\Phi_{II}}{ds}\Big|_{s=s_2^-}\approx\frac{N^{5/3}}{2^{2/3}}$,
whilst $N^{3/2} \frac{d\Psi_{III}}{ds}\Big|_{s=s_2^+}
\approx\frac{N^{5/3}}{4\, 2^{2/3}}$ for large $N$.  At the transition
point $s=s_2$, there is also a change of concavity of the curve: the
rate function in regime {\bf II} is convex ($
\frac{d^2\Phi_{II}}{ds^2}>0$ for all $s<9/4$) and has a minimum at
$s=\bar{s}=2$, whereas the rate function in regime {\bf III} is
concave ($ \frac{d^2\Psi_{III}}{ds^2}<0$ for all $s>2$).
\\

\textbf{Scaling $\Sigma_2=S\approx O(1)$
and limit $S\rightarrow 1$ (unentangled state)}
\\

In the far-right tail of the distribution $\Sigma_2=S\approx O(1)$
($S\gg s/N$, $S\leq 1$) and the maximal eigenvalue $t\approx O(1)$
whereas $\zeta$ (and all the other eigenvalues) remain of order
$O(1/N)$.  In this limit, equations \eqref{eq:t} and \eqref{eq:zeta}
become:
\begin{equation}\label{RegIIIlimS1}
  S\approx t^2\;\; {\rm and} \;\; L\approx 4 (1-t)
  \;\; {\rm as} \; t\approx O(1)\,.
\end{equation}
The saddle point energy  in Eq. \eqref{EnergyIII} reduces to:
$
E_S\left[\rho_c,t\right]\approx -\frac{N^2}{2} \ln\left(1-t\right)
+ N^2\left(
\frac{\ln N}{2}+\frac{1}{4}\right)-N \ln N +O(N)$
as  $S\approx O(1)$
 with $t=\sqrt{S}$.
Using  Eq. \eqref{NormIII}, we get an explicit expression
for the rate function
$\Phi(N,S)=\left( E_S\left[\rho_c,t_c \right]-
    E\left[\rho^*,t^* \right] \right)/N^2$
for large $N$:
\begin{equation}\label{PhiIIIq2limS1}
\Phi(N,S)\approx
 \frac{\left(E_S\left[\rho_c,t\right] - N^2\left(
\frac{\ln N}{2}+\frac{1}{4}\right)\right)}{N^2}
\approx -\frac{1}{2} \ln\left(1-\sqrt{S}\right)\equiv
\Phi_{III}(S)\,.
\end{equation}
We conclude that
\begin{equation}\label{PdfSigma2limS1}
P\left(\Sigma_2=S,N\right)\approx e^{-\beta  N^2 \Phi_{III}(S)}
\approx \left(1-\sqrt{S}\right)^{\frac{\beta N^2}{2}}
\;\;\textrm{for large $N$, fixed $S$}\,.
\end{equation}
The difference of scaling with respect to regimes {\bf I} and {\bf II}
comes from the scaling of $\Sigma_2$: in regimes {\bf I} (resp. {\bf
  II}), we had $\Phi(N,s/N)\rightarrow \Phi_I(s)$
(resp. $\Phi_{II}(s)$) for large $N$, whereas here we have:
$\Phi(N,S)\rightarrow \Phi_{III}(S)$ for large $N$ and fixed $S\approx
O(1)$.  As $S=s/N$ with fixed $s$ and large $N$, which corresponds to
the limit $S\rightarrow 0$ in this scaling, we find $ N^2
\Phi_{III}(S)\approx N^{3/2}\,\sqrt{s}/2$ which is also the limit
$s\rightarrow \infty$ of $N^{3/2} \Psi_{III}(s)$.  The right tail
(where $S\approx O(1/N)$) and the far-right tail (where $S\approx
O(1)$) of the distribution match smoothly.

As $\Sigma_2=S$ tends to its maximal value $1$, the maximal eigenvalue
$t\rightarrow 1$ and $L\rightarrow 0$. At $S=1$, only one eigenvalue,
the maximal one $\lambda_{max}=t$, is nonzero (and equal to one). This
corresponds to an unentangled state (situation (i)).  The probability
of an unentangled state (i.e. $\Sigma_2 \rightarrow 1$) is thus
vanishingly small for large $N$.
\\

\begin{figure}
\includegraphics[width=9cm]{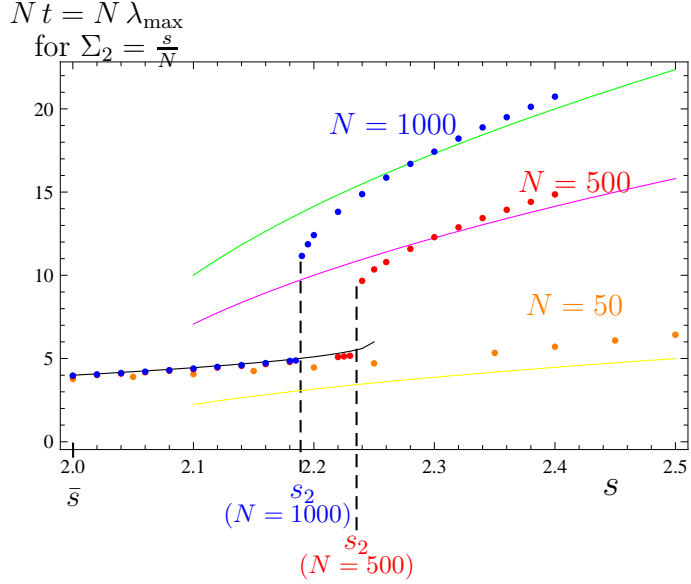}
\caption{Maximal eigenvalue $\lambda_{\rm max}=t$ corresponding to a
  fixed value of the purity $\Sigma_2=s/N$ plotted against $s$ for
  different values of $N$.  Analytical predictions (solid lines) are
  compared with numerical simulations (points\,: Monte Carlo data,
  method 2 with density).  The theory predicts for large $N$ a sudden
  jump of $t$ from a value $t\approx \zeta= L(s)/N$ with
  $L(s)=2(3-\sqrt{9-4 s})$ (within regime {\bf II}, $s<s_2$) to a
  much larger value $t\approx \frac{\sqrt{s-2}}{\sqrt{N}}$ (regime
  {\bf III}, $s>s_2$). We clearly see this jump in numerical
  simulations for $N=500$ at $s_2\approx 2.23$ and $N=1000$ at
  $s_2\approx 2.18$.  For $N=50$, finite-size corrections to
the large $N$ asymptotics are considerable enough to smear the
jump in $t$. Because of the choice of
  scaling on the plot, $t N$ as a function of $s$, the plots of the
  maximal eigenvalue in regime {\bf II} are expected to be the same
  for different $N$ (for large $N$), whereas the plots for regime {\bf
    III} differ by a factor $\sqrt{N}$.  }\label{fig:max1}
\end{figure}


\subsubsection{General $q>1$}

Using again Tricomi's theorem and imposing the constraints $\int
\rho_c=1$ and $\rho_c(\zeta)$, we find that the optimal charge density
for the $N-1$ smallest eigenvalues is given by:
\begin{equation}\label{RhocIIIgen}
  \rho_c(\lambda)=\frac{1}{\pi (N-1)}\sqrt{\frac{\zeta-\lambda}{\lambda}}\,
  \left[ A + B\:
    \,_2F_1\left(1,1-q,\frac{3}{2},1-\frac{\lambda}{\zeta}\right)
    +\frac{C}{t-\lambda}
  \right]\,,
\end{equation}
where $A=\mu_1$, $B=\mu_2 2 q \zeta^{q-1}
\frac{\Gamma(q+1/2)}{\sqrt{\pi} \Gamma(q)}$ and
$C=\sqrt{\frac{t}{t-\zeta}}$ and $\,_2F_1$ is a hypergeometric
function $\,_2F_1(a,b,c,z)=\sum_{n=0}^{\infty} \frac{(a)_n
  (b)_n}{(c)_n} \frac{z^n}{n!}$, with $(a)_n=a (a+1)...(a+n-1)$
denoting the raising factorial (Pochhammer symbol).  The Lagrange
multipliers $\mu_1$ and $\mu_2$ are given by:
\begin{eqnarray}
\mu_1\!\!\!&=&\!\!\! \frac{4}{(q-1)\zeta^2}\left[q N \zeta -2(q+1)
  +\sqrt{\frac{t}{t-\zeta}}
\left\{(2 q +2)t-(2 q +1)\zeta  \right\}\right]\,,\nonumber\\
\mu_2\!\!\!&=&\!\!\! \frac{\sqrt{\pi} \Gamma(q+2)}{\zeta^{q+1}\Gamma(q+1/2)q(q-1)
}\left[4- N \zeta 
  +\sqrt{\frac{t}{t-\zeta}}
\left\{3 \zeta -4 t  \right\}\right]\,, \label{mu1mu2regIII}
\end{eqnarray}
where $\zeta$ and $t$ are solutions of the following system of
equations:
\begin{eqnarray}
&(a)&\!\!\!\! S-t^q = \frac{\zeta^{q-1}\Gamma(q+1/2)}{\sqrt{\pi} \Gamma(q+1)}
\left\{ 2-\frac{N \zeta}{2}\left(\frac{q-1}{q+1}\right)
+\sqrt{\frac{t}{t-\zeta}}\left[
\zeta\left( \frac{3 q+1}{2q +2}\right)\right.\right. \nonumber\\
&& \;\;\;\;  \;\;\;\; \; \left.\left. -2 t
\right]
 \right\}
+ \sqrt{\frac{t}{t-\zeta}}\:
\frac{\zeta^{q+1} \Gamma(q+1/2)}{2 t \sqrt{\pi} \Gamma(q+2)}
\; _2F_1\left(1,q+\frac{1}{2},2+q,\frac{\zeta}{t}\right)\,,
\label{eq:tQ}\\
&(b)&\!\!\!\!\!\mu_1 \sqrt{\frac{t- \zeta}{t}}\:+
q \mu_2 t^{q-1}=\frac{\zeta}{2 t (t-\zeta)}+\nonumber \\
&&\mu_2 \frac{\zeta^q  \Gamma(q+\frac{1}{2})}
{t\sqrt{\pi}\Gamma(q)}
\, _2F_1\left(1,q,q+1,\frac{\zeta}{t}\right)\,,\hspace{1cm}
\label{eq:zetaQ}
\end{eqnarray}
with $\mu_1=\mu_1(\zeta,t)$ and $\mu_2=\mu_2(\zeta,t)$ given in
Eq. \eqref{mu1mu2regIII}.

These equations can be solved analytically for large $N$
and the solutions are qualitatively the same as for $q=2$.

For $S=N^{1-q}\, s$ with $\bar{s}(q)<s<s_0(q)$ (where
$s_0(q)=\frac{\Gamma(q+1/2)}{2 \sqrt{\pi} \Gamma(q+2)} \left(\frac{4
    (1+q)}{q}\right)^q$, see regime {\bf II}), there exist two
different solutions for the pair $(\zeta,t)$.  The first solution is
of the form $t\approx \zeta$ with $\zeta \approx O(1/N)$. This is
exactly (to leading order in $N$) the solution of regime {\bf II} (see
below, ``first solution'').  There is also a second solution with $t
\gg \zeta$, more precisely $\zeta =L/N$ with $L\sim O(1)$ and
$t\approx O(1/N^{1-1/q})$ for $S\approx N^{1-q}\, s$, and
$\zeta\approx O(1/N)$ (see below, ``second solution'').  For $s$ close
to $\bar{s}(q)$, the first solution dominates (regime {\bf II}), but
at some point $s=s_2(q,N)> \bar{s}(q)$ given in Eq. \eqref{s2qgen1},
the second solution, with $t \gg \zeta$, starts to dominate (its
energy becomes lower): this is regime {\bf III}.

For $S>N^{1-q}\,s_0(q)$, i.e. $s>s_0$, only the second solution
remains: the upper bound of the density support scales as $\zeta =L/N$
with $L\sim O(1)$ while the maximal eigenvalue is much larger than all
other eigenvalues: $t \gg \zeta$.

In both cases (as for $q=2$), for large $N$ the upper bound $\zeta$
remains of order $\sim O(1/N)$ ($\zeta\sim \lambda_{\rm typ}$).  We
shall thus write $\zeta=\frac{L}{N}$ with $L\sim O(1)$.  On the other
hand (as for $q=2$), the maximal eigenvalue $t$ scales from $O(1/N)$
(as $S\rightarrow N^{1-q}\bar{s}(q)$) to $O(1)$ (as $S\rightarrow
1^-$).
\\

\textbf{Scaling $S=N^{1-q}\, s$ with $s\sim O(1)$ : first solution
$t\approx\zeta$ with $\zeta \sim O(1/N)$ (regime {\bf II})}
\\

For $S=N^{1-q}\, s$ with $s\sim O(1)$ for large $N$, the solution of
regime {\bf II} still exists as long as $s<s_0(q)$.  We recover this
solution from the Eq.  \eqref{eq:tQ} and \eqref{eq:zetaQ} with the
scaling $t=\frac{T}{N}$ and $\zeta=\frac{L}{N}$ with $T\approx L \sim
O(1)$, where the maximal eigenvalue $t$ remains very close to the
other eigenvalues ($t\approx\zeta$ for large $N$), it does not play a
special role.  Using Eq. \eqref{NormIII}, we finally get
$\Phi(N,s/N)=\left(E_S[\rho_c,t]-E_S[\rho_c,t]\Big|_{s=2}\right)/N^2 =
\Phi_{II}(s)$, the same expression as in Eq. \eqref{PhiII} of regime
{\bf II}.

However, for $ s > \bar{s}(q)$ there exists a second solution that
becomes energetically more favorable at some point $s_2(q,N)$.
Therefore regime {\bf II} is only valid for $s_1<s<s_2$.
\\

\textbf{Scaling $S=N^{1-q}\, s$ with $s\sim O(1)$ : second solution
$t\gg\zeta$ (regime {\bf III})}
\\

For $S=N^{1-q}\, s$ with $s>\bar{s}(q)$, there exists a second
solution where one eigenvalue ($\lambda_{\rm max}=t$) becomes much
larger than the other eigenvalues : $t \gg \zeta$.  In this limit,
Eq. \eqref{eq:tQ} and \eqref{eq:zetaQ} give for large $N$:
\begin{eqnarray}\label{eq:tZetasol2Q}
 t\approx\frac{\left[s-\bar{s}(q)\right]^{1/q}}{N^{1-1/q}}
\;\; \;{\rm and}\;\;\;
\zeta \approx \frac{4}{N}
 \left[1-\left\{\frac{s- \bar{s}(q)(1+q)/2}{\left[s-\bar{s}(q)\right]^{1-1/q}}\right\}
\, \frac{1}{N^{1-1/q}}\right]\,.
\end{eqnarray}
For $S\rightarrow 1$, which implies $s\rightarrow \infty$ as
$N\rightarrow \infty$, we find $t\approx s^{1/q}\,N^{1/q-1}=S^{1/q}$
and $\zeta \approx\frac{4}{N}(1-t)\;$.

We can compute the saddle point energy in this limit replacing $t$ and
$\zeta$ by the expressions given in Eq. \eqref{eq:tZetasol2Q} for
large $N$.  Finally, we get $N^2
\Phi(N,s/N)=E_{S}\left[\rho_c,t\right]- N^2\left(\frac{\ln
    N}{2}+\frac{1}{4} \right)\approx
N^{1+\frac{1}{q}}\,\frac{\left[s-\bar{s}(q)\right]^{1/q}}{2}$ for
large $N$ (see Eq. \eqref{NormIII}) and the pdf of $\Sigma_q$ is thus
given for large $N$ by:
\begin{equation}\label{PdfSigmaqRegIIILargeN1}
P\left(\Sigma_q=N^{1-q}\, s,N \right)\approx \exp\left\{-\beta
  N^{1+\frac{1}{q}
}\,   \Psi_{III}(s)\right\}\,,
\end{equation}
where
\begin{equation}\label{PsiIIIgen1}
\Psi_{III}(s)=\frac{\left[s-\bar{s}(q)\right]^{1/q}}{2}\;\;\; \textrm{for large $N$}\,.
\end{equation}

The solution of regime {\bf III} becomes energetically more favorable,
that is $ N^{1+\frac{1}{q}} \Psi_{III}(s)< N^{2} \Phi_{II}(s)$, at some
point $s_2(q,N)$ defined by $ N^{1+\frac{1}{q}} \, \Psi_{III}(s_2)=
N^{2} \Phi_{II}(s_2)$.  Therefore
\begin{equation}\label{s2qgen1}
  s_2(q,N)\approx  \bar{s}(q)+\frac{\left[\sqrt{q/2} \, 
    (q-1) \, \bar{s}(q) \right]^{2
    q/(2 q-1)}}{
  N^{(q-1)/(2 q-1)}}\;
  \;\textrm{for large $N$}\,.
\end{equation}
At $s=s_2$, there is an abrupt transition from regime {\bf II} to {\bf
  III}.  The maximal eigenvalue $t$ jumps from a value $t\approx
\frac{T}{N}$ with $T\sim O(1)$ and $t$ very close to $\zeta$ to a
value $t\approx \frac{\left[s-\bar{s}(q)\right]^{1/q}}{N^{1-1/q}}$
much larger than the other eigenvalues ($t \gg \zeta$).  The rate
function $\Phi(N,s/N)$ given by
\begin{equation}\label{RateFunQregIIandIII1}
N^2 \, \Phi(N,s/N)=\left\{\begin{array}{ll}
N^{2}\,  \Phi_{II}(s) &{\rm for} \;\; s<s_2\,,\\
N^{1+\frac{1}{q}}\, \Psi_{III}(s) &{\rm for} \;\; s>s_2 \,,
\end{array}\right.
\end{equation}
is continuous but its derivative is discontinuous. For large $N$, we
have indeed $N^2\, \frac{d\Phi}{ds}\big|_{s_2^-}\approx N^{\frac{3
    q-1}{2 q-1}}\, \left\{ (q-1)\sqrt{q/2}\: \bar{s}(q)
\right\}^{\frac{2-2 q}{2 q-1}} $, whilst $
\frac{d\Phi}{ds}\big|_{s_2^+} \approx \frac{d\Phi}{ds}\big|_{s_2^-}
/(2 q)$.  At the transition point $s=s_2$, there is also a change of
concavity of the curve: the rate function in regime {\bf II} is convex
($ \frac{d^2\Phi_{II}}{ds^2}>0$) and has a minimum at $s=\bar{s}$,
whereas the rate function in regime {\bf III} is concave ($
\frac{d^2\Psi_{III}}{ds^2}<0$).

\section{Distribution of the Renyi entropy $S_q$}
\label{sec:Renyi}

In section \ref{sec:PdfSigmaq}, we have computed the full distribution
of $\Sigma_q=\sum_{i=1}^N \lambda_i^q$ for large $N$.  A simple change
of variable gives the distribution of the Renyi entropy
$S_q=\frac{1}{1-q} \ln \left[\sum_i \lambda_i^q\right] =\frac{1}{1-q}
\ln \left[\Sigma_q\right]$.  The scaling $\Sigma_q=N^{1-q} s$ for
large $N$ implies $S_q =\ln N -\frac{\ln s}{q-1}$. This means that
typical values of $S_q$ will be of order $S_q\approx \ln N -z$ with
$z\approx O(1)$ for large $N$.  The parameter $z=\frac{\ln s}{q-1}$ is
nonnegative and its minimum $z=0$ corresponds to $S_q=\ln N$, which
corresponds to the maximally entangled state.

The distribution of the entropy is thus given for large $N$ by:
\begin{equation}\label{PdfEntropyQ}
P\left(S_q = \ln N-z \right)\approx
\left\{\begin{array}{ll}
\exp\left\{ -\beta N^2\: \phi_{I}(z)\right\}&{\rm for} \;\; 0<z\leq z_1(q) \,,\\
& \\
\exp\left\{-\beta N^2\: \phi_{II}(z)\right\}&{\rm for} \;\; z_1(q)<z\leq z_2(q) \,,\\
& \\
\exp\left\{-\beta N^{1+\frac{1}{q}}\; \psi_{III}(z)\right\}&{\rm for} \;\; z>z_2(q) \,.
\end{array}
\right.
\end{equation}
The three regimes are the same as for $\Sigma_q$.  The rate functions
$\phi_I$, $\phi_{II}$ and $\psi_{III}$ are simply obtained from the
rate functions $\Phi_I$, $\Phi_{II}$ and $\Psi_{III}$ for the
distribution of $\Sigma_q$ (see Eq. \eqref{PdfPurSummary}) by the change
of variable $s=\exp\left[(q-1)z\right]$, e.g.
$\phi_I(z)=\Phi_I\left(e^{(q-1)z}\right)$.
Explicit expressions of the functions
$\Phi_I$ and $\Phi_{II}$ are given in Eq. \eqref{PhiIq2} and
\eqref{PhiIIq2} for $q=2$, and in Eq.  \eqref{PhiII} for a
general $q>1$; an explicit expression of $\Psi_{III}$ is given in
Eq. \eqref{PsiIIIgen} for a general $q>1$ (and in Eq.~\eqref{PsiIIIq2}
for $q=2$).

The critical points are given by
\begin{equation}\label{CritPointsEntropy}
z_1(q)=\frac{\ln s_1(q)}{q-1}\;\;{\rm and}\;\;
z_2(q,N)=\frac{\ln s_2(q,N)}{q-1}\,,
\end{equation}
where $s_1$ and $s_2$ are the critical points for $\Sigma_q$
(see Eqs.~\eqref{CritPoints} and \eqref{CritPointFiniteN}).

The distribution of the entropy $S_q$ has the same qualitative
behaviour as that of $\Sigma_q$ : it is a highly peaked distribution
with Gaussian behaviour around the mean value and non-Gaussian tails.
Again, the average value of $S_q$ coincides with the most probable
value for large $N$, $\langle S_q \rangle\approx \ln N -\bar{z}(q)$
where $\bar{z}(q)$ is the minimum of $\phi_{II}$:
\begin{equation}\label{AverEntropy}
\langle S_q \rangle\approx \ln N -\bar{z}(q)\;\;
{\rm with }\;\; \bar{z}(q)=\frac{\ln \bar{s}(q)}{q-1} =
\frac{1}{q-1}\, \ln \left[\frac{\Gamma(q+1/2)}{ \Gamma(q+2)}
\, \frac{4^q}{\sqrt{\pi}} \right]\,.
\end{equation}
The rate function $\phi_{II}(z)$ has a quadratic behaviour around 
$z=\bar{z}(q)$:
$\phi_{II}(z)\approx \frac{(z-\bar{z}(q))^2}{q}$.
Therefore, the distribution of the entropy $S_q$
has a Gaussian behaviour around its average:
\begin{equation}\label{EntropyGauss}
P\left(S_q = \ln N-z \right)\approx \exp\left\{
-\beta N^2 \frac{(z-\bar{z}(q))^2}{q}\right\}\;\; \textrm{for}\;\;
z\approx \bar{z}(q) \,,
\end{equation}
which gives the variance of the distribution:
\begin{equation}\label{VarEntropy}
{\rm Var} \,S_q\approx\frac{q}{2 \beta N^2} \;\;\textrm{for large $N$}\,.
\end{equation}

\subsection{Limit $q\rightarrow 1^+$ : von Neumann entropy}

As $q\rightarrow 1^+$, the Renyi entropy $S_q$ tends to the von
Neumann entropy $S_{\rm VN}=-\sum_i \lambda_i \ln \lambda_i$.  The
limit $q\rightarrow 1$ is singular for the distribution of
$\Sigma_q$\,: because of the constraint $\Sigma_1=\sum_i \lambda_i=1$,
the distribution tends to a Dirac-$\delta$ function. The variance
tends to zero ($\sigma_q^2 \rightarrow 0$) and the mean value
$\bar{s}(q)$ as well as the critical point $s_1(q)$ and $s_2(q)$ tend
to $1$.  However, due to the factor $1/(1-q)$ in the definition of
$S_q$, the limit $q\rightarrow 1$ is not at all singular for the
entropy $S_q$. Taking this limit only requires to be careful.  For
$S_{\rm VN}$ (as for $S_q$ for $q>1$), there are three regimes in the
distribution:
\begin{equation}\label{PdfSVN}
P\left(S_{\rm VN} = \ln N-z \right)\approx
\left\{\begin{array}{ll}
\exp\left\{ -\beta N^2\: \phi_{I}(z)\right\}&{\rm for} \;\; 0<z\leq z_1 \,,\\
& \\
\exp\left\{-\beta N^2\: \phi_{II}(z)\right\}&{\rm for} \;\; z_1<z\leq z_2 \,,\\
& \\
\exp\left\{-\beta \frac{N^2}{\ln N}\; \phi_{III}(z)\right\}&{\rm for} \;\; z>z_2 \,,
\end{array}
\right.
\end{equation}
where $\phi_{II}$ and $\phi_{III}$ are respectively given in
Eqs.~\eqref{PhiIISVN} and \eqref{PsiIIISVN}.  For $q\rightarrow 1$, we
get: $\bar{z}(q)=\frac{\ln \bar{s}(q)}{q-1} \rightarrow 1/2$ (where
$\bar{z}(q)$ is given in Eq. \eqref{AverEntropy}).  We thus recover
the already known mean value of the von Neumann entropy (see \cite{don
  page}) in the case $c=1$ ($M\approx N$):
\begin{equation}\label{AverSVN}
\langle S_{\rm VN} \rangle \approx \ln N -\frac{1}{2} \;\;
\textrm{for large $N$}\,.
\end{equation}
The critical points separating the three regimes are given by (limit
$q\rightarrow 1$ in Eqs.~\eqref{CritPointsEntropy} and
\eqref{CritPoints}):
\begin{equation}\label{CritPointsSVN}
z_1 = \frac{2}{3} -\ln \frac{3}{2}\approx 0.26 \;\; {\rm and}\;\;
z_2\approx\bar{z}=\frac{1}{2}\,.
\end{equation}

We easily obtain the expression of the rate function $\phi_{II}$ in
regime {\bf II} by taking the limit $q\rightarrow 1$.  We get:
\begin{equation}\label{PhiIISVN}
  \phi_{II}(z)=
  -\frac{1}{2} \ln \left(\frac{L}{4}\right)
  + \frac{8}{ L^2}-\frac{6}{ L}+1 \,,
\end{equation}
where $L=L(z)$ is the solution of (limit $q\rightarrow 1$ in
Eq. \eqref{eq:L})
\begin{equation}\label{eq:LSVN}
\ln\left(\frac{L}{4}\right) -\frac{L}{8}+1=z \,.
\end{equation}
For large $N$, the mean value corresponds to the minimum of
$\phi_{II}$.  The quadratic approximation of $\phi_{II}$ around this
minimum $z\approx\bar{z}$ gives the Gaussian behaviour of the pdf of
$S_{\rm VN}$ around its average and thus the variance in the large $N$
limit:
\begin{equation}\label{VarSVN}
  \langle S_{\rm VN} \rangle \approx \ln N -\bar{z}\;\; {\rm with}
  \;\; \bar{z}=\frac{1}{2}\;\;{\rm and}\; \;{\rm Var}\, S_{\rm VN} \approx\frac{1}{2 \beta N^2}\,.
\end{equation}

The limit $q\rightarrow 1$ for the regime {\bf III} is a bit more
subtle. We would expect the rate function to be of the form $N^2
\psi_{III}(z)$, but $\psi_{III}=\Psi_{III}(e^{(q-1)z})$ (in
Eq. \eqref{PsiIIIgen}) vanishes as $q\rightarrow 1$.  The rate
function actually scales as $N^2/\ln N$ (rather than $N^2$ as one could
na\"ively expect).  This can be shown by a more detailed analysis of the
equations \eqref{eq:tQ} and \eqref{eq:zetaQ} in the limit
$q\rightarrow 1$.  The solution $t \gg \zeta$ is actually given for
$q\rightarrow 1$ by:
\begin{equation}\label{tZetaSVN}
t\approx\frac{z-1/2}{\ln N}\;\;{\rm and}\;\;
\zeta\approx \frac{4}{N}\left( 1+\frac{1-z}{\ln N}\right) \,.
\end{equation}
The saddle point energy can be computed in this limit.
We finally find:
\begin{equation}\label{PsiIIISVN}
-\ln P(S_{\rm VN}=\ln N-z)\approx \beta \frac{N^2}{\ln N}(z-1/2)\,;\quad
\phi_{III}(z)=z-\frac{1}{2}\,.
\end{equation}

\subsection{Limit $q\rightarrow \infty$ : maximal eigenvalue}

As $q\rightarrow \infty$ the Renyi entropy $S_q$ tends to $-\ln
\lambda_{\rm max}$ where $ \lambda_{\rm max}$ is the maximal
eigenvalue.  Again, the limit is singular for the distribution of
$\Sigma_q$ but not for $S_q$. There are the same three regimes in the
distribution of $\lambda_{\rm max}$ for large $N$ as in the
distribution of the Renyi entropy.

For large $N$, the typical scaling is $S_q \approx \ln N -z$, thus
$-\ln \lambda_{\rm max} \approx \ln N -z$ or $\lambda_{\rm
  max}\approx \frac{e^z}{N}$.  Setting $t=e^z$, we have
$\lambda_{\rm max}=t/N$.  In particular, the mean value is
given by $\bar{t}/N$ where $\bar{t}=\lim_{q\rightarrow \infty}
\exp(\bar{z}(q)) =\lim_{q\rightarrow \infty} \left[\bar{s}(q)\right]^{
  \frac{1}{q-1}}=4$, implying
\begin{equation}\label{AverLambdaMax}
\langle \lambda_{\rm max} \rangle \approx \frac{4}{N}\,.
\end{equation}
The first critical point is $t_1= \lim_{q\rightarrow \infty}
\left[s_1(q)\right]^{ \frac{1}{q-1}}=4/3$.  The second critical point
is $t_2=\bar{t}=4$.  The three regimes in the distribution of the
maximal eigenvalue are the following:
\begin{equation}\label{PdfSumLambdaMax}
P\left(\lambda_{\rm max}=\frac{t}{N}\right)\approx
\left\{\begin{array}{ll}
e^{ -\beta N^2 \chi_{I}(t)}&{\rm for} \;\; 1<t\leq 4/3 \;\; 
\textrm{(reg. {\bf I})}\,,\\
& \\
e^{-\beta N^2 \chi_{II}(t)}&{\rm for} \;\; 4/3<t\leq 4 \;\;
\textrm{(reg. {\bf II})}\,,\\
& \\
e^{-\beta N \chi_{III}(t)}&{\rm for} \;\; t>4\;\; \textrm{(reg. {\bf III})}\,.
\end{array}
\right.
\end{equation}
The rate functions can be explicitely computed.  The rate function in
regime {\bf I} is given by:
\begin{equation}\label{PhiILambdaMax}
\chi_I(t)=-\frac{1}{2} \ln(t-1)\;\;\;\; {\rm for}\;\;
1<t\leq 4/3 \,.
\end{equation}
In regime {\bf II}, we find:
\begin{equation}\label{PhiIILambdaMax}
\chi_{II}(t)=4\frac{(1-t)}{t^2}-\frac{1}{2}\ln\left(\frac{t}{4}\right)
+\frac{3}{4}\;\;\;\; {\rm for}\;\;
4/3<t\leq 4 \,.
\end{equation}
Finally, in regime {\bf III} the maximal eigenvalue detaches from the
sea of the other eigenvalues and we get:
\begin{equation}\label{PsiIIILambdaMax}
\chi_{III}(t)=\frac{\sqrt{t(t-4)}}{2}-2 \ln (\sqrt{t}+\sqrt{t-4})
+2 \ln2\;\;\;\;\textrm{for $t>4$}\,.
\end{equation} 
Again, at the first critical point $t_1=4/3$, the rate function $\chi$
is continuous and twice differentiable, but its third derivative is
discontinuous: $\frac{d^3 \chi_{I}}{dt^3}=-27$ but $\frac{d^3
  \chi_{II}}{dt^3}=-999/64$.  The average value $\bar{t}=4$ is the
minimum of $\chi_{II}$.  At the second critical point $t_2=4$, the
rate function is continuous but not differentiable.

Exactly as we did for $\Sigma_q$, we can also consider the regime
where $\lambda_{\rm max}=T$ ($T \gg t/N$): the far-right tail of the
distribution.  We find:
\begin{equation}\label{PsiIIIplusLambdaMax}
P\left(\lambda_{\rm max}=T\right)\approx e^{-\beta N^2 \chi_+(T)}\;\;
\chi_+(T)=-\frac{1}{2} \ln(1-T)\;\; {\rm for}\;\; 0<T<1 \,,
\end{equation}
which matches smoothly regime {\bf III}.  We have indeed: $N
\chi_{III}(t)\approx N \frac{t}{2}$ as $t\rightarrow \infty$ and $N^2
\chi_+(t) \approx N^2 \frac{T}{2}\approx N \frac{t}{2}$ as
$T\rightarrow 0$ with $T=t/N$.
\\

\textbf{Ideas of proof}
\\

Regimes {\bf II} and {\bf III} can be derived by taking carefully the
limit $q\rightarrow \infty$ (directly in the expression of the rate
function for regime {\bf II} but more carefully for regime {\bf III}).
The distribution of $\lambda_{\rm max}$ can also be computed directly
(without taking the limit $q\rightarrow \infty$).  This gives the same
results for regimes {\bf II} and {\bf III} and gives also an explicit
expression for regime {\bf I} (where the rate function is not
explicitely known for a general $q>1$).  We can actually calculate the
cumulative distribution ${\rm Prob}\left(\lambda_{\rm max }\leq Z
\right)$ by the same Coulomb gas method as before.  This is indeed
easier to compute because the probability that $\lambda_{\rm max} \leq
Z$ is the probability that all the eigenvalues $\lambda_i$ are smaller
than $Z$.  We can thus compute this probability with the Coulomb gas
method, with a continuous density $\rho(x)=1/N \, \sum_i
\delta(x-\lambda_i N)$ and with the constraint that no eigenvalue
exceeds $Z$:
\begin{equation}\label{CdfLambdaMax}
P\left( \lambda_{\rm max} \leq Z \right)
\propto \int \mathcal{D}\rho \, e^{-\beta N^2 E_Z\left[\rho\right]}\,.
\end{equation}
The energy reads
\begin{eqnarray}
E_Z\left[\rho \right]&=&-\frac{1}{2}\int_0^{Z}\int_{0}^Z \, \rho(x)
\rho(x') \ln|x-x'|\, dx\, dx'
+\mu_0 \left(\int_0^Z \rho(x) dx -1\right)\nonumber\\
&&+\mu_1  \left(\int_0^Z x \rho(x) dx -1\right) \,,
\label{EnLambdaMax}
\end{eqnarray}
where the Lagrange multipliers $\mu_0$ and $\mu_1$ enforce the two
constraints $\int \rho=1$ (normalization of the density) and $\int x
\rho =1$ (unit sum of the eigenvalues: $\sum_i \lambda_i=1$).  The
saddle point method gives:
\begin{equation}\label{SaddleLambdaMax}
P\left( \lambda_{\rm max} \leq Z \right)
\propto  \, e^{-\beta N^2 E_Z\left[\rho_c\right]}\,,
\end{equation}
where $\rho_c$ minimizes the effective energy $E_Z$.  This yields
regimes {\bf I} and {\bf II}.  Exactly as for $S_q$, in regime {\bf
  III}, the maximal eigenvalue detaches from the sea of the other
charges (eigenvalues), it must be taken into account separately from
the continuous density of the other eigenvalues.

In regime {\bf I}, the optimal charge density has a finite support
$[L_1,L_2]$ and vanishes at $L_{1,2}$ (exactly as for $\Sigma_q$). We
get the rate function $\chi_I$ in Eq. \eqref{PhiILambdaMax}.

In regime {\bf II}, the optimal charge density has a finite support
$]0,L]$, vanishes at $L$ but diverges at the origin with a square root
divergence (exactly as for $\Sigma_q$). We get the rate function
$\chi_{II}$ in Eq. \eqref{PhiIILambdaMax}.  This expression can also
be obtained by taking the limit $q\rightarrow \infty$ of the
expression in Eq. \eqref{PhiII} of $\Phi_{II}$, valid for a
general $q$ (for $\Sigma_q$).

In regime {\bf III}, the maximal eigenvalue is much larger than the
others and we get $\chi_{III}$ in Eq. \eqref{PsiIIILambdaMax}.  The
limit $q\rightarrow \infty$ in the rate function $\psi_{III}$ for a
general $q$ gives: $\psi_{III} \longrightarrow t/2$.  This is actually
equal to $\chi_{III}(t)$ only in the limit $t\rightarrow \infty$, but
not for all $t>4$. For $q>1$, regime {\bf III} is characterized by
$t\approx T/N^{1-\frac{1}{q}} \gg \zeta$ as $\zeta\approx L/N$, which
becomes $t\approx T/N >\zeta$ in the limit $q\rightarrow \infty$. The
maximal eigenvalue is larger than the other eigenvalues, but not much
larger.  We cannot anymore assume $t \gg \zeta$ in the computation of
the energy.  We must compute carefully the energy $E_S\left[\rho_c,t\right]$
 in this limit.
For this computation, we use the complete expression of $E_S$:
for $q=2$, this expression was given in Eq. \eqref{EnergyIII};
for a general $q$, we have a similar but more complicated expression.
We use this expression in the limit where $t$ and $\zeta$ are both of order
one (with $t>\zeta$) and where $q\rightarrow \infty$.
We finally get $\chi_{III}(t)$ as given in Eq. \eqref{PsiIIILambdaMax}.
\\

\subsubsection{Typical fluctuations around the average: Tracy-Widom 
distribution}

We have seen that the average value of the maximal eigenvalue, in the
large $N$ limit, is given by $\langle \lambda_{\rm max}\rangle \approx 4/N$.
Of course, $\lambda_{\rm max}$ fluctuates around this average from
sample to sample. The Coulomb gas method presented in this subsection
captures fluctuations $\sim O(1/N)$ around this mean, i.e., large
fluctuations that are of the same order of magnitude as the mean
itself. We have seen that the probability of such large $\sim O(1/N)$
fluctuations is very small, indicating that they are rare atypical
fluctuations. The typical fluctuations around the mean occur at a
much finer scale around this mean which is not captured by
the Coulomb gas method.

To compute the distribution of such typical fluctuations, we
start from the joint distribution in \eqref{jpdfEV}.
The cumulative probability of the maximum can be written as the multiple integral
\begin{equation}
P\left(\lambda_{\rm max}\le 
Z\right)\propto \int_0^Z\ldots\int_0^Z P(\lambda_1,\lambda_2,\ldots, 
\lambda_N)d\lambda_1\, d\lambda_2\ldots d\lambda_N
\label{max1}
\end{equation}
Next we can replace the delta function $\delta \left(\sum_{i=1}^N 
\lambda_i-1\right)$ by its integral representation: 
$\delta(x)=(1/{2\pi i})\int dp e^{px}$ where the 
integral runs over the imaginary axis. This gives, for $M=N$,
\begin{equation}
P\left(\lambda_{\rm max}\le
Z\right)\propto \int \frac{dp}{2\pi i} e^p \int_{[0,Z]} 
\left[\prod_{i=1}^N d\lambda_i\right] 
e^{-p\sum_{i=1}^N \lambda_i}\, \prod_{i=1}^N 
\lambda_i^{\frac{\beta}{2} -1}\:
\prod_{i<j} |\lambda_i -\lambda_j|^{\beta}.
\label{max2}
\end{equation}
Rescaling $\lambda_i\to (\beta/{2p})\lambda_i$, one can recast the integral
as
\begin{equation}
P\left(\lambda_{\rm max}\le
Z\right)\propto \int_{-i\infty}^{i\infty} \frac{dp}{2\pi i} e^p\, p^{-\beta 
N^2/2}\, 
\int_{[0,2pZ/{\beta}]}
\left[\prod_{i=1}^N d\lambda_i\right]
e^{-\frac{\beta}{2}\sum_{i=1}^N \lambda_i}\, \prod_{i=1}^N
\lambda_i^{\frac{\beta}{2} -1}\:
\prod_{i<j} |\lambda_i -\lambda_j|^{\beta}.
\label{max3}
\end{equation}
The integral over $\lambda_i$'s is just proportional to the cumulative 
distribution of the maximum of the Wishart matrix, i.e., the
$P^{Wishart}\left(\lambda_{\rm max}\le 2pZ/{\beta}\right)$.
This latter quantity, in the large $N$ limit, is known~\cite{Johansson,Johnstone} 
to converge
to a limiting distribution known as 
the Tracy-Widom distribution~\cite{TW}, i.e,
\begin{equation}
P^{Wishart}\left(\lambda_{\rm max}\le y\right)\to 
F_{\beta}\left[\frac{(y-4N)}{2^{4/3}N^{1/3}}\right]
\label{maxtw}
\end{equation}
where $F_{\beta}(x)$ satisfies a nonlinear differential equation~\cite{TW}. 
Using this result in \eqref{max3}, we get, in the large $N$ limit,
\begin{equation}
P\left(\lambda_{\rm max}\le
Z\right)\propto \int_{-i\infty}^{i\infty} \frac{dp}{2\pi i} e^{p-\frac{\beta}{2} 
N^2 \log(p)}\, F_{\beta}\left[\frac{\frac{2p}{\beta}Z-4N}{2^{4/3} 
N^{1/3}}\right].
\label{max4}
\end{equation}
The integral over $p$ can now be evaluated via the saddle point method. To
leading order for large $N$, one can show that the saddle point
occurs at $p^*= \beta N^2/2$ that just minimise the exponential
factor $e^{p-\frac{\beta}{2}
N^2 \log(p)}$. Hence, to leading order in large $N$, we obtain our main result
\begin{equation}
P\left(\lambda_{\rm max}\le
Z\right)\approx F_{\beta}\left[\frac{Z-4/N}{2^{4/3}N^{-5/3}}\right].
\label{max5}
\end{equation} 

This shows that $\lambda_{\rm max}$ in our problem typically fluctuates
on a scale $O(N^{-5/3})$ around its average $4/N$,  
\begin{equation}\label{TW}
{\rm typical}\;\; \lambda_{\rm max}=\frac{4}{N}+2^{4/3}N^{-5/3} \chi_{\beta} \,,
\end{equation}
where the distribution of the random variable $\chi_{\beta}$ is the
Tracy-Widom probability density function $g_{\beta}(x)=dF_{\beta}(x)/dx$.  
Around 
the
mean value we have then
\begin{equation}\label{PdfTW}
P\left(\lambda_{\rm max}=\frac{t}{N}\right)\approx
N^{5/3}\, g_{\beta}\left(2^{-4/3} N^{2/3} (t-4)\right) \,.
\end{equation}\\

\noindent {\bf Matching between the tails of the Tracy-Widom distribution
and the large deviation rate functions}\\ 

For Gaussian and Wishart matrices, it has been recently 
demonstrated~\cite{DM,vivo1,MV}
that the Tracy-Widom density describing the probability of typical fluctuations 
of the 
largest eigenvalue matches smoothly, near its tails,
with the left and right rate functions that describe the probabilty of
atypical large fluctuations. It would be interesting to see if
the same matching happens in our problem as well. Indeed, we find
that the tails of the Tracy-Widom distribution match smoothly to
our previously obtained rate
functions. 

For the left tail of the Tracy-Widom density, it is known~\cite{TW}
that $g_{\beta}(x)\sim
\exp\left\{-\frac{\beta}{24}|x|^3 \right\}$ for $x\rightarrow
-\infty$.  Therefore $P\left(\lambda_{\rm max}=\frac{t}{N}\right)\sim
\exp\left\{-\beta N^2 \frac{|t-4|^3}{384}\right\}$.
On the other hand, for the rate function to the left of the mean
describing large fluctuations of $\sim O(1/N)$ is given
in \eqref{PhiIILambdaMax}. Taking the limit $t\to 4^-$, we find
$\chi_{II}(t)\approx -
\frac{(t-4)^3}{384} $ thus matching smoothly with the left tail
of the Tracy-Widom density.

For the right tail, one knows~\cite{TW} $g_{\beta}(x)\sim \exp\left\{-\frac{2
    \beta}{3}\, x^{3/2} \right\}$ for $x\rightarrow +\infty$.
Therefore $P\left(\lambda_{\rm max}=\frac{t}{N}\right)\sim
\exp\left\{-\beta N \frac{(t-4)^{3/2}}{6}\right\}$.
On the other hand, the rate function describing large
fluctuations of order $\sim O(1/N)$ to the right of the mean
is given in \eqref{PsiIIILambdaMax}. Expanding to leading order
for  $t\rightarrow 4^+$, we get: $\chi_{III}(t)\approx
\frac{(t-4)^{3/2}}{6} $  which clearly matches smoothly to
the right tail of the Tracy-Widom density.


\section{Numerical simulations}
\label{sec:MonteCarlo}

To verify the analytical predictions derived in the preceding
sections, we simulated the joint distribution of eigenvalues in
Eq. \eqref{jpdfEV}:
\begin{eqnarray}\label{jpdfEV1}
P(\lambda_1,\ldots,\lambda_N)&=& B_{M,N}\,
\delta\left( \sum_i \lambda_i-1 \right) \;
\prod_{i=1}^N \lambda_i^{\frac{\beta}{2} (M-N+1) -1}\: 
\prod_{i<j} |\lambda_i -\lambda_j|^{\beta}\nonumber\\
&=&B_{M,N}\:\delta\big( \sum_i \lambda_i-1 \big) \;
e^{-\beta E\left[\{\lambda_i\} \right]}\,,
\end{eqnarray}
where the effective energy $E\left[{\lambda_i}\right]$ is given by
Eq. \eqref{jpdfEeff}.  We sampled this probability distribution using
a Monte Carlo Metropolis algorithm (see \cite{Krauth}).

\subsection{Standard Metropolis algorithm}

We start with an initial configuration of the $\lambda_i$'s satisfying
$\sum_{i=1}^N \lambda_i=1$ and $\lambda_i>0$ for all $i$.  At each
step, a small modification $\{\lambda_i\} \longrightarrow
\{\lambda_i'\}$ is proposed in the configuration space.  In our
algorithm, the proposed move consists of picking at random a pair
$(\lambda_j,\lambda_k)$ (with $j\neq k$) and proposing to modify them
as $(\lambda_j,\lambda_k) \longrightarrow (\lambda_j +\epsilon,
\lambda_k -\epsilon)$, which naturally conserves the sum of the
eigenvalues. $\epsilon$ is a real number drawn from a Gaussian
distribution with mean zero and with a variance that is set to achieve
an average rejection rate $1/2$.

The move is rejected if one of the eigenvalues becomes negative.
Otherwise, the move is accepted with the standard probability
\begin{equation}
\label{DetailedBalance}
p=\min\left(\frac{
P(\lambda_1',...,\lambda_N')}{
P(\lambda_1,...,\lambda_N)},1 \right)
=\min\left( e^{-\beta \left(E\left[ \{\lambda_i' \}\right]-
E\left[\{\lambda_i \}\right]\right)},1\right)\,,
 \end{equation}
 and rejected with probability $1-p$.  This dynamics enforces detailed
 balance and ensures that at long times the algorithm reaches thermal
 equilibrium (at inverse ``temperature'' $\beta$) with the correct
 Boltzmann weight $e^{-\beta E\left[\{\lambda_i\}\right]}$ and with
 $\sum_i \lambda_i=1$.

 At long times (from about $10^6$ steps in our case), the Metropolis
 algorithm thus generates samples of $\{\lambda_i\}$ drawn from the
 joint distribution in Eq. \eqref{jpdfEV1}. We can then start to
 compute some functions of the $\lambda_i$'s, e.g. the purity
 $\Sigma_2=\sum_i \lambda_i^2$, and construct histograms, e.g. for the
 density, the purity, etc..

 However, as the distribution of the purity (as well as the one of the
 eigenvalues) is highly peaked around its average, a standard
 Metropolis algorithm does not allow to explore in a ``reasonable'' time
 a wide range of values of the purity.  The probability to reach a
 value $\Sigma_2=s/N$ decreases rapidly with $N$ as $e^{-\beta
   N^2 \Phi(s)}$ where $\Phi(s)$ is a positive constant (for $s$
 different from the mean value\,: $s\neq \bar{s}$).  Therefore, we
 modified the algorithm in order to explore the full distribution of
 the purity and to compare it with our analytical predictions.

\subsection{Method 1 : Conditional probabilities}

It is difficult to reach large values $\Sigma_2=s/N$ of the purity
($s>\bar{s}$).  The idea is thus to force the algorithm to explore the
region $s\geq s_c$ for different values of $s_c$. We thus add in the
algorithm the constraint $s\geq s_c$.  More precisely, we start with
an initial configuration that, in addition to $\sum_i \lambda_i=1$ and
$\lambda_i>0$ for all $i$, satisfies also $\sum_i\lambda_i^2\geq
s_c/N$. At each step, the proposed move is rejected if $\sum_i
\lambda_i'^2 < s_c/N$.  If $\sum_i \lambda_i'^2 \geq s_c/N$, then the
move is accepted or rejected exactly with the same Metropolis rules as
before.  Because of the new constraint $s\geq s_c$, the moves are
rejected much more often than before. Therefore the variance of the
Gaussian distribution $P(\epsilon)$ has to be taken smaller to achieve
a rejection rate $1/2$.

We run the program for several values of $s_c$ (about $20$ different
values) and we construct a histogram of the purity for each value
$s_c$.  This gives the conditional probability distribution
$P\left(\Sigma_2=\frac{s}{N} \big| \Sigma_2 \geq \frac{s_c}{N}
\right)$.  Again, as the distribution of the purity is highly peaked,
the algorithm can only explore a very small range of values of $s$ -
even for a large running time (about $10^8$ steps).  The difference
with the previous algorithm is that we can now explore small regions
of the form $s_c \leq s \leq s_c + \eta$ for every $s_c$, whereas before 
we could only explore the neighbourhood of the mean value
$\bar{s}$.

The distribution of the purity is given by
\begin{equation}
P\left(\Sigma_2=\frac{s}{N}\right)=
P\left(\Sigma_2=\frac{s}{N} \big| \Sigma_2 \geq \frac{s_c}{N} \right)*
P\left(\Sigma_2\geq \frac{s_c}{N}\right)\;\;\textrm{(for $s_c<s$)}\,.
\end{equation}
Therefore the rate function reads:
\begin{eqnarray}
  \Phi(s)&=&-\frac{1}{\beta\, N^2}\ln P\left(\Sigma_2=\frac{s}{N}\right)
  \nonumber \\
  &=&-\frac{1}{\beta \, N^2} 
  \left[ \ln P\left(\Sigma_2=\frac{s}{N} \big| \Sigma_2 \geq \frac{s_c}{N} \right)
    + \ln P\left(\Sigma_2\geq \frac{s_c}{N}\right) \right] \,.
\end{eqnarray}
The histogram constructed by the algorithm with the constraint $s\geq
s_c$ is the rate function $\Phi_{s_c}(s)=-\frac{1}{\beta \, N^2} \ln
P\left(\Sigma_2=\frac{s}{N} \big| \Sigma_2 \geq \frac{s_c}{N}
\right)$.  $\Phi_{s_c}(s)$ differs from the exact rate function
$\Phi(s)$ by an additive constant that depends on $s_c$.  In order to
get rid of this constant, we construct from the histogram giving
$\Phi_{s_c}(s)$ the derivative $\frac{d\Phi_{s_c}(s)}{ds}$.  This
derivative is equal to $\frac{d\Phi(s)}{ds}$ and the constants disappear.
We can now compare numerical data with the derivative of the
analytical expression for the rate function $\Phi(s)$.

We can also come back to $\Phi(s)$ from its derivative using an
interpolation of the data for the derivative and a numerical
integration of the interpolation.  This allows to compare directly the
numerical results with the theoretical rate function $\Phi(s)$.

We can follow the same steps to explore the region on the left of the
mean value $s<\bar{s}$ by adding in the simulations the condition
$\sum_i \lambda_i^2 \leq \frac{s_c}{N}$ (instead of $\sum_i
\lambda_i^2 \geq \frac{s_c}{N}$) for several values of $s_c <
\bar{s}$.

We typically run the simulations for $N=50$ and $10^8$ iterations. As
figure \ref{fig:N50Meth1} shows, numerical data and analytical
predictions agree very well for regimes {\bf I} and {\bf II} (rate
functions given in Eqs.~\eqref{PhiIq2} and \eqref{PhiIIq2}).
For regime {\bf III}, finite-size effects are important and agreement
holds for large but finite $N$ analytical formulae (taking as rate
function the expression of the energy in Eq.~\eqref{EnergyIII} with
$t$ and $\zeta$ numerical solutions of the system of equations
\eqref{eq:t} and \eqref{eq:zeta}). The agreement would degrade for the
asymptotic rate function giving only the dominant term for very large
$N$ ( Eq.~\eqref{PsiIIIq2}).  Finite-size effects are also important
for the transition between regimes {\bf II} and {\bf III}.  Large-$N$
data are crucial to see clearly this abrupt transition with a sudden
jump of the maximal eigenvalue.  For $N=50$, the transition appears
indeed to be smoothed out. This observation can be rationalized as
follows. At the transition ($s=s_2 $), the maximal eigenvalue $t$ is
expected to jump for large $N$ from a value $\sim \frac{5}{N}$ to a
much larger value $\sim \sqrt{\frac{s-2}{N}}$, yet for $N=50$ we have
$\frac{5}{N}> \sqrt{\frac{s-2}{N}}$ for all $s<9/4$. We thus conclude
that no jump can be seen at $N=50$ and much larger $N$ are needed.
Adapting the simulation method to cope with this challenge is the
subject of the next subsection.

\subsection{Method 2 : Simulation of the density of eigenvalues
(and conditional probabilities)}

We want to be able to run simulation for very large values of $N$.
The idea is to simulate the density
$\rho(\lambda)=\frac{1}{N}\sum_i\delta(\lambda-\lambda_i)$ rather than
the eigenvalues themselves.  In the previous scheme, a configuration was
made of $N$ variables, the $N$ eigenvalues.  In the new code, we have
$k+2 \ll N$ variables:

(1) the maximal eigenvalue $t$.

(2) the upper bound of the density support $\zeta$ ($\zeta <t$).

(3) the value of the density at each point $x_i = \frac{i \zeta}{k}$
(for $0\leq i <k$).

We must enforce the condition $\rho(\zeta)=0$, i.e. $\rho(x_k)=0$ by
definition of the upper bound $\zeta$ of the density support.  The
idea is to replace the real density by a linear approximation of the
density defined by its value at $x_i$ for $0\leq i\leq k$.

These $k+2$ variables describing the maximal eigenvalue and the
density of the other eigenvalues simulate configurations with $N \gg
k$ eigenvalues, for example $N=1000$ with $k=50$. The number of
eigenvalues $N$ appears in the expression of the energy (and in the
constraints).  With this new code, we can now simulate configurations
with many eigenvalues in a reasonable time.
\\

\textbf{The algorithm}
\\

From the analytical calculations, we expect that the density diverges
when $\lambda\rightarrow 0^+$ as $\rho(\lambda)\sim
\frac{1}{\sqrt{\lambda}}$.  In order to get a better approximation in
our code, we choose to discretize a regularized form of the density
$\bar{\rho}(\lambda)\equiv \sqrt{\lambda}\rho(\lambda)$.  Our $(k+2)$
variables are thus:

(1) the maximal eigenvalue $t$.

(2) the upper bound of the regularized density support $\zeta$ ($\zeta
<t$), which is the same as the upper bound of the density support.

(3) the value of the regularized density at each point $x_i = \frac{i
  \zeta}{k}$ (for $0\leq i <k$): $z_i\equiv \bar{\rho}(x_i)$.

In the Monte Carlo simulation, we compute the energy as well as the
constraints ($\sum_i\lambda_i=1$, etc.) by using a linear
interpolation of the regularized density $\bar{\rho}(\lambda)$\,:
\begin{equation}
 \tilde{\rho}(\lambda)=z_i +\frac{z_{i+1}-z_i}{x_{i+1}-x_i}\:(\lambda-x_i)
\;\;{\rm for}\;\; \lambda \in [x_i,x_{i+1}[\,,
\end{equation}
with $z_i=\bar{\rho}(x_i)$ (in particular $z_k=0$).  Integrals such as
$\int d\lambda \, \lambda \, \rho(\lambda)$ are computed using the
linear interpolation as\,:
\begin{equation}
  \int_0^{\zeta} d\lambda\rho(\lambda)\lambda
  \approx \frac{4}{15} \left( \frac{\zeta}{k}\right)^{\frac{3}{2}} \left[
    z_0+\sum_{i=1}^{k-1} z_i \left\{ (i+1)^{\frac{5}{2}}+(i-1)^{\frac{5}{2}}-2\, i^{\frac{5}{2}} \right\}\right]\,.
\end{equation}

There are two constraints for the density\,: the normalization $\int
\rho =1$ and the unit sum of the eigenvalues $t+(N-1)\int \lambda
\rho=1$. We start from an initial configuration satisfying these
constraints\,: for example, we can take for the initial $\rho$ a
density of the form of the (normalized) average density
$\rho(\lambda)=\frac{2}{\pi \zeta}\sqrt{
  \frac{\zeta-\lambda}{\lambda}}$ and fix $t$ with the unit sum
constraint $t=-(N-1)\int \lambda \rho+1$.  Initially, we also choose
$\zeta$ not too large such that the condition $\sum_i \lambda_i^2
>s_c/N$ is satisfied (for a fixed value of $s_c$), exactly as in the
code with conditional probabilities.

At each step, we propose a move in the configuration space (our $k+2$
variables) that naturally enforces the two constraints $\int \rho =1$
and $t+(N-1)\int \lambda \rho=1$ (unit sum).  More precisely, at each
step we choose randomly three integers between $0$ and $k+1$\,:
$i_1<i_2<i_3$.
\begin{itemize}
\item If $i_3<k$ (case 1), we propose a move
  $(z_{i_1},z_{i_2},z_{i_3})\longrightarrow (z_{i1}+\alpha_1 \epsilon,
  z_{i2}+\alpha_2 \epsilon, z_{i3}+\alpha_3 \epsilon)$, where
  $\epsilon$ is drawn from a Gaussian distribution with zero mean and
  a variance adjusted to have the standard rejection rate $1/2$ at the
  end. $\alpha_1$, $\alpha_2$ and $\alpha_3$ are constants that are
  chosen such that the constraints $\int \rho =1$ and $t+(N-1)\int
  \lambda \rho=1$ (unit sum of eigenvalues) are satisfied:

$\alpha_1 = \left[(i_3+1)^{3/2}+(i_3-1)^{3/2}-2 i_3^{3/2}\right]
 \left[(i_2+1)^{5/2}+(i_2-1)^{5/2}-2 i_2^{5/2}\right]
$
\\
$-
 \left[(i_2+1)^{3/2}+(i_2-1)^{3/2}-2 i_2^{3/2}\right]
\left[(i_3+1)^{5/2}+(i_3-1)^{5/2}-2 i_3^{5/2}\right]$
\\
$\alpha_2$ and $\alpha_3$ are obtained from $\alpha_1$
by cyclic permutation of $i_1$, $i_2$ and $i_3$.
\item If $i_1<i_2<i_3=k$ (case 2), we propose a move
  $(\zeta,z_{i_1},z_{i_2})\longrightarrow (\zeta+\epsilon,z_{i1}+
  \epsilon_1, z_{i2}+ \epsilon_2)$ where $\epsilon$ is drawn from a
  Gaussian distribution with zero mean and a variance adjusted to have
  the standard rejection rate $1/2$ at the end (different from the
  variance of case 1), and where $\epsilon_1$ and $\epsilon_2$ are
  functions of $\epsilon$, $i_1$ and $i_2$ fixed by the two
  constraints ($\int\rho=1$ and unit sum).
\item If $i_1<i_2<k$ and $i_3=k+1$ (case 3), we propose a move
  $(t,z_{i_1},z_{i_2})\longrightarrow (t+\epsilon,z_{i1}+ \epsilon_1,
  z_{i2}+ \epsilon_2)$, where, exactly as in case 2, $\epsilon$ is
  drawn from a Gaussian distribution, and $\epsilon_1$ and
  $\epsilon_2$ are functions of $\epsilon$, $i_1$ and $i_2$ fixed by
  the two constraints ($\int\rho=1$ and unit sum).
\item If $i_1<i_2=k$ and $i_3=k+1$ (case 4), we propose a move
  $(\zeta,z_{i_1},t)\longrightarrow (\zeta+\epsilon,z_{i1}+
  \epsilon_1, t+dt)$, where $\epsilon$ is drawn from a Gaussian
  distribution (same as in case 2), and $\epsilon_1$ and $dt$ are
  functions of $\epsilon$ and $i_1$ fixed by the two constraints
  ($\int\rho=1$ and unit sum).
\end{itemize}

Then, if $\zeta >t$, if $\zeta<0$, if $z_i<0$ or if $\sum_i
\lambda_i^2 <s_c/N$, that is $(N-1)\int \lambda^2 \, \rho +t^2<s_c/N$,
the move is rejected. Otherwise we compute the energy of the new
configuration $E_{\rm new}$ and accept the move with the usual
Metropolis probability $p=\min\left( e^{-\beta \left(E_{\rm new}- E
    \right)},1\right)$ (and reject it with probability $1-p$).

Direct inspection of the previous rules shows that detailed balance is
satified. Therefore, after a large number of iterations, thermal
equilibrium with the appropriate Boltzmann weight is reached and we
can start to construct histograms of the density and the purity.  We
verified that for $N=50$ (simulated with $k+2$ variables, where
$k=20$) we recover the results of the direct Monte Carlo (where we
simulate directly the eigenvalues).  For $N=500$ and $N=1000$ (with
$k=50$), we get very interesting results that can be used to test the
large-$N$ analytical predictions (see Eqs.~\eqref{PhiIq2} and
\eqref{PhiIIq2} for regimes {\bf I} and {\bf II} and
Eq. \eqref{PsiIIIq2} for regime {\bf III}): figure
\ref{fig:N1000Meth2} shows the good agreement between theory and
numerical simulations with this second method, for the distribution of
the purity $\Sigma_2=\sum_i\lambda_i^2$ with $N=1000$.  As figure
\ref{fig:max1} shows, we can really see the abrupt jump of the maximal
eigenvalue and the change of behaviour of the rate function
(discontinuous derivative), which is expected at the transition
between regime {\bf II} and regime {\bf III} for very large $N$.

The simulations also provide solid support to the fact that a single
eigenvalue detaches from the sea in regime III. One might indeed
wonder whether configurations with multiple charges detaching from the
sea could be more favorable. This was ruled out by measuring the area
of the rightmost ``bump'' in the density of charges (see
Fig.~\ref{fig:schemadens}) and verifying that it corresponds to a
single charge. This fact is also intuitively rationalized as
follows. Let us consider configurations with two charges, $\lambda_1$
and $\lambda_2$ ($\lambda_1\geq \lambda_2$), detaching from the
sea. As in Eq.~(\ref{EeffSqIII}), we require
$\lambda_1^q+\lambda_2^q=t^q$ and we consider the quantity ${\cal
  C}=1-\lambda_1-\lambda_2$, which quantifies the compression of the
sea of charges and would replace $1-t$ in the $\mu_1$ constraint in
Eq.~(\ref{EeffSqIII}). The smaller is ${\cal C}$, the stronger is the
compression of the sea (with the other constraints remaining the
same).  Since the charges repel each other, the energy of the
configuration is expected to increase as ${\cal C}$ gets smaller. An
elementary calculation shows that, due to the convexity of $\lambda^q$
for $q>1$, ${\cal C}$ is minimum when $\lambda_1=\lambda_2=2^{-1/q}t$
while its maximum (minimum energy) is attained at the boundary
$\lambda_1=t$, $\lambda_2=0$, corresponding indeed to a single charge
detaching from the sea.

\section{Conclusion}

In this paper, by using a Coulomb gas method, we have computed the
distribution of the Renyi entropy $S_q$ for $q>1$ for a random pure
state in a large bipartite quantum system, i.e. with a large dimension
$N$ of the smaller subsystem.  We have showed that there are three
regimes in the distribution $P\left(S_q=\ln N -z\right)$ that are a
direct consequence of two phase transitions in the associated Coulomb
gas.

(i) Regime {\bf I} corresponds to the left tail of the distribution
($0<z<z_1(q)$).
In this phase, the effective potential
seen by the Coulomb charges
has a minimum at a nonzero point. 
The charge density has a finite support over $[L_1,L_2]$
(and vanishes at $L_1$ and $L_2$), the charges
accumulate around the minimum of the potential.

(ii) Regime {\bf II} describes the central part of the distribution
($z_1(q)<z<z_2(q)$), and in particular the vicinity of the mean value
$\bar{z}(q)$.  At the transition between regimes {\bf I} and {\bf II},
the third derivative of the rate function (logarithm of the
distribution) is discontinuous.  In this phase, the charges
concentrate around the origin, the charge density has a finite support
over $[0,L]$ with a square-root divergence at the origin.  Close to
the mean value of $S_q$, the distribution is Gaussian.

(iii) Regime {\bf III} describes the right tail of the distribution
($z>z_2(q)$),
corresponding to a more and more unentangled state.
In this phase, one charge splits off the sea of the other charges.
The transition between regimes {\bf II} and {\bf III} is abrupt
with a sudden jump of the rightmost charge (largest eigenvalue).
There is thus a discontinuity of the derivative of the rate function
and the scaling with $N$ changes at this point.

A by-product of our results is the fact that, although the average
entropy is close to its maximal value $\ln N$, the probability of a
maximally entangled state is actually very small.  The probability
density function of the entropy indeed vanishes at $z=0$ (far left
tail), i.e. at $S_q=\ln N$, which is the maximally entangled
situation.  Similar properties and three different regimes are also
obtained in the limit $q\rightarrow 1$, which gives us the
distribution of the von Neumann entropy, and in the limit
$q\rightarrow \infty$, which yields the distribution of the maximal
eigenvalue.

\subsubsection*{Acknowledgements}
We thank Sebastien Leurent for useful discussions.
\\

\medskip

{\it Note:} Soon after we submitted our first paper (the short version
published in \cite{prlBE}), an independent work appeared in the Arxiv
(\textit{arXiv:0911.3888}) (now published in
~\cite{pasquale}) where the phase transitions in the distribution of
the purity (the case $q=2$) are also discussed, but with a slightly
different point of view (the Laplace transform of the distribution is
studied).


\end{document}